\documentclass[aps,twocolumn,pra,superscriptaddress,10pt]{revtex4-2}

\usepackage[T1]{fontenc}
\usepackage{dsfont}
\usepackage{amsmath}

\usepackage{bm}
\usepackage{amssymb}
\usepackage{graphicx}
\usepackage{braket}
\usepackage{subfigure}
\usepackage{comment}
\usepackage{xcolor}
\usepackage{oplotsymbl}
\usepackage{lmodern}
\usepackage{empheq}

\usepackage[colorlinks=true,urlcolor=blue,citecolor=blue,linkcolor=blue]{hyperref}
\usepackage{float}											

\newcommand{\ie}{\hbox{\em i.e.{}}}
\newcommand{\eg}{\hbox{\em e.g.{}}}

\newcommand{\lhs}{\hbox{l.h.s.{}}}
\newcommand{\rhs}{\hbox{r.h.s.{}}}

\newcommand{\SUt}{\text{SU}(2)}
\newcommand{\un}{\mathfrak{u}(n)}
\newcommand{\td}{\text{d}}
\def \Tr {\text{Tr}}

\newcommand{\diff}{\mathop{}\!\mathrm{d}}

\newcommand*\xbar[1]{%
  \hbox{%
  \kern.1em
    \vbox{%
      \hrule height 0.2pt 
      \kern0.25ex
      \hbox{%
        \kern-0.1em
        \ensuremath{\scriptstyle #1}%
        \kern-0.05em
      }%
    }%
  }%
} 

\newcommand{\tetrahedron}{
  \mathchoice
    {\includegraphics[height=1.4ex]{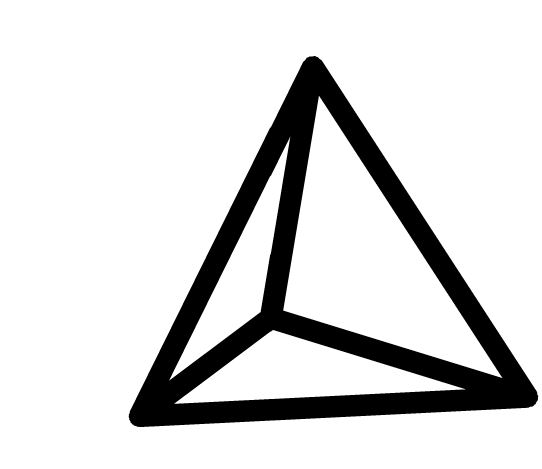}} 
    {\includegraphics[height=1.4ex]{tetraImage}} 
    {\includegraphics[height=1.2ex]{tetraImage}} 
    {\includegraphics[height=.6ex]{tetraImage}} 
}

\newcommand{\trbyp}{
  \mathchoice
    {\includegraphics[height=1.4ex]{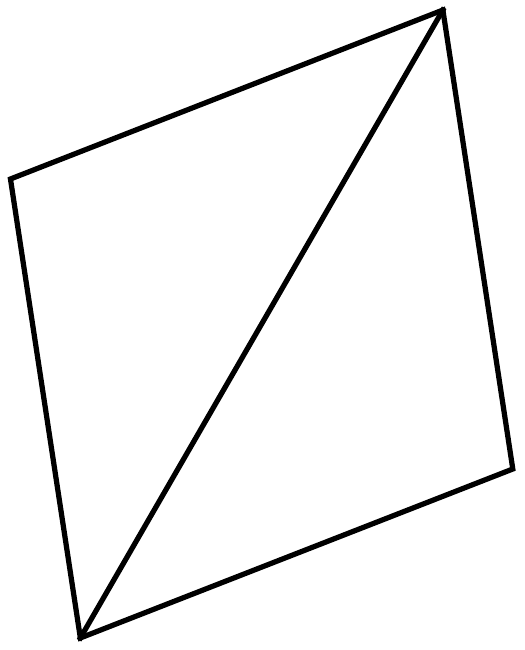}} 
    {\includegraphics[height=1.4ex]{triangularDipyramid}} 
    {\includegraphics[height=1.2ex]{triangularDipyramid}} 
    {\includegraphics[height=.6ex]{triangularDipyramid}} 
}

\newcommand{\ssssquare}{{\scriptscriptstyle \square}}
\makeatletter
\g@addto@macro\bfseries{\boldmath}
\makeatother
\newcommand{\rema}[1]{\textcolor{black}{#1}}
\newcommand{\remal}[1]{\textcolor{black}{#1}}
\begin{document} 
\title{Universality in fidelity-based quantum metrology}
\author{Luis Arag\'on-Mu\~noz}
\email{luis.aragon@correo.nucleares.unam.mx}
\affiliation{Instituto de Ciencias Nucleares\\ Universidad Nacional Autónoma de México PO Box 70-543, 04510, CDMX, México}
\author{Chryssomalis Chryssomalakos}
\email{chryss@nucleares.unam.mx}
\affiliation{Instituto de Ciencias Nucleares\\ Universidad Nacional Autónoma de México PO Box 70-543, 04510, CDMX, México}
\author{Ana Gabriela Flores-Delgado}
\email{ana.flores@correo.nucleares.unam.mx}
\affiliation{Instituto de Ciencias Nucleares\\ Universidad Nacional Autónoma de México PO Box 70-543, 04510, CDMX, México}
\author{John Martin}
\email{jmartin@uliege.be}
\affiliation{Institut de Physique Nucléaire, Atomique et de Spectroscopie, CESAM, University of Liège
\\
B-4000 Liège, Belgium}
\author{Eduardo Serrano-Ens\'astiga}
\email{ed.ensastiga@uliege.be}
\affiliation{Institut de Physique Nucléaire, Atomique et de Spectroscopie, CESAM, University of Liège
\\
B-4000 Liège, Belgium}
\date{\today}
\begin{abstract}
We consider the problem of identifying the quantum spin states that are the optimal sensors of a given transformation, averaged over all possible orientations of the spin system. Our geometric approach to the problem is based on a fidelity criterion and is entirely general, encompassing \rema{any} unitary transformation
. This formalism leads to a universality result: For any value of the spin, there exists a zero-measure subset of states that can be the optimal sensors for certain transformations and the worst sensors for others, and this set does not depend on the transformation under consideration. In other words, some spin states are simply the best (or worst) sensors, regardless of what they detect.
\end{abstract}
\maketitle
%
\section{Introduction}
\label{Introduction}
Quantum-enhanced metrology seeks to overcome the fundamental sensitivity limits of classical devices by employing quantum systems as probes~\cite{RevModPhys.90.035005,doi:10.2514/1.J062707,PhysRevLett.132.190001}. Platforms such as Bose–Einstein condensates, photonic systems, and cold-atom ensembles have already been used successfully in experiments as quantum sensors, to detect, for instance, magnetic fields or rotations~\cite{PhysRevLett.111.143001,mao2023quantum,Ferretti:24,Bou.etal:17,yang_minute-scale_2025}. To describe this class of transformations in simple terms, we focus only on single-spin systems, with a corresponding finite-dimensional Hilbert space --- other physical platforms, such as multiphotonic systems, can be shown to be equivalent to these using the correspondence provided by the Schwinger representation~\cite{Agarwal_2012}. We consider transformations of such systems, implemented by multiplication of the spinor $\ket{\psi}$ that describes the spin state by a matrix $M$. In the simplest case, $M$ might model an interaction with an external magnetic field \rema{or} a physical rotation, the common denominator being the presence of a particular direction in space (that of the magnetic field or the rotation axis). In many practical cases, this direction is not known a priori, since transformations are not always controlled and may result, \emph{e.g.}, from fluctuating external fields. Or, one just does not know the orientation of the probe state, as in searches for new interactions in fundamental physics~\cite{PhysRevA.108.010101}. In such cases, a natural approach is to look for a quantum state whose sensitivity is optimal on average over all possible field orientations. A physical example is a homogeneous magnetic field, where the orientation is determined by a single axis (the field direction). These scenarios have been extensively studied, both when the direction of the axis is known~\footnote{This case is equivalent to phase estimation~\cite{RevModPhys.90.035005,PhysRevA.57.4736}.}, or when sensitivity is desired for several directions simultaneously,  including the case where the direction is averaged uniformly over the sphere~\cite{Kolenderski_Demkowicz-Dobrzanski_2008,ZGoldberg_2021,Piotrak2024,PhysRevA.95.052125,Mar.Wei.Gir:20,PhysRevA.111.022435,Czartowski2024,du2025characterizing}. Experimental realizations have been demonstrated across several platforms~\cite{PhysRevLett.104.133601,Bou.etal:17,Goldberg_2025,Ferretti:24}. Related protocols also include using random states drawn from different statistical ensembles~\cite{Goldberg_2025} and employing randomized measurements~\cite{s27y-gbrp}. Recent studies further examine the role of self-consistent many-body dynamics in bosonic atomic ensembles for metrological tasks~\cite{PhysRevLett.132.240803}.

In more general settings, the transformation matrix $M$ cannot be encoded by a single vector specifying a direction in space, but instead depends on more complex information. For example, when the spin couples to the gradient of an electric field $\mathbf{E}$, the relevant object is the tensor $\partial E^i/\partial x^j$, which, when symmetric (in Cartesian coordinates) and nondegenerate, defines the orthonormal frame of its eigenvectors --- it is this frame (or the related SO(3) matrix) that encodes the orientation of ``the fields''. The simplest interactions of this type  are Hamiltonians quadratic in the spin operators. This includes quadrupolar electric interactions or Hamiltonians capable of generating squeezings~\cite{PhysRevLett.131.183003,DeMille2024,MA201189}. Experimental implementations of these sensors appear in tests of fundamental physics, for instance, in the search for an electric dipole moment for the electron~\cite{PhysRevLett.131.183003}, in the detection of new interactions between dark matter and probe-state spins~\cite{Blanchard2023}, or in tests of local Lorentz-symmetry violation~\cite{PhysRevLett.120.103202,PhysRevA.99.042118}, to name just a few (see e.g.\ Ref.~\cite{PhysRevA.108.010101} for a survey).

In this work, we develop a geometric approach to the determination of the pure spin states that are optimal in detecting a given transformation, in a rotationally averaged sense. The figure of merit we use is the SU(2)-average fidelity: for each state $\rho$, we evaluate the transition probability between the initial state and the transformed one, and then average this quantity over the entire SU(2)-orbit of the state. \rema{This fidelity criterion can be viewed as a symmetry-averaged analogue of the Loschmidt echo, obtained by averaging the fidelity over the SU(2) orbit of the probe state. Here, the unitary operator $V$ plays the role of the relative evolution between unperturbed and perturbed dynamics~\cite{Gorin2006}. In this language, optimal sensors are precisely those states that minimize the averaged echo, \ie, those whose dynamics is most sensitive to the applied perturbation.} We declare a state to be an optimal sensor, for the given transformation, if it minimizes the above average fidelity over all pure spin states of the same spin.  This fidelity-based metrology problem has already been studied in the case where the transformation is a rotation~\cite{PhysRevA.95.052125,Kolenderski_Demkowicz-Dobrzanski_2008,Mar.Wei.Gir:20,PhysRevA.111.022435}, leading to highly counterintuitive results: as the rotation angle varies, optimal sensors change discontinuously at certain critical rotation angles, the number and values of which depend on the spin quantum number $s$~\cite{Mar.Wei.Gir:20,PhysRevA.95.052125}. To make things even more intriguing, the states that are optimal sensors in a certain interval of the rotation angle can be the worst sensors for another interval, and vice versa~\cite{PhysRevA.95.052125,Mar.Wei.Gir:20}. 

The novelty in our approach is that we present a systematic geometric method to calculate the optimal quantum sensor for any kind of transformation with respect to this average fidelity. 
The approach is quite general, allowing us to study several physical scenarios simultaneously, such as those mentioned above, including \rema{any} unitary \rema{transformation}, \emph{e.g.}, rotations and squeezings. It is remarkable that for small spin quantum numbers, the set of optimal states (for almost all transformations) contains only a finite number of them, in other words, only a handful of quantum states act as universal optimal sensors of arbitrary spin transformations. 

This paper is organized as follows: Section~\ref{SubSec.Metro} \rema{first states the problem. We then} provide a short review of the mathematical and physical concepts used throughout the work in Section~\ref{MathPre}. Section~\ref{Sec.Ofbs} contains the general formalism for \rema{arbitrary} transformations and for the evaluation of $\SUt$-average fidelities, \rema{both in the infinitesimal and finite cases}. Then, in Section~\ref{Sec.Gen.Uni}, we present several explicit examples,  involving rotations and squeezings, highlighting the potential of our approach. We conclude with some final remarks in Section~\ref{Sec.Conclusions}. Readers interested only in the final results (and not their derivation) can consult \rema{Section~\ref{SubSec.Metro}} and then proceed to Section~\ref{Sec.Gen.Uni} for a summary of the proposed approach and its applications.
\section{Statement of the problem}
\label{SubSec.Metro}
\subsection{\rema{Optimal sensing of unitary transformations}}
\label{SubSec.OptimalSensing}
We consider the pure quantum states of a spin-$s$ system, represented by vectors in the Hilbert space $\mathcal{H} \cong \mathbb{C}^n$, $n=2s+1$. Vectors that differ by a complex factor represent physically equivalent states, so that the physical state space is the complex projective space $\mathbb{C}P^{n-1}$, \ie, the space of rays in $\mathcal{H}$. Thus, a state $\ket{\psi} \in \mathcal{H}$ gets projected to a point $[\psi] \in \mathbb{C}P^{n-1}$, and the latter may be represented by an $n$-dimensional Hermitian (density) matrix $\rho=\ket{\psi}\bra{\psi}$, satisfying, additionally,
\begin{equation}
    \label{rhoprop}
    \rho^2=\rho
    \, ,
    \quad
    \Tr \rho=1
    \, .
\end{equation}
We are interested in quantifying how distinguishable (from itself) a spin state becomes after a given transformation is applied to it. Starting from an initial state $\rho$, the system transforms under
\begin{equation}
    V(\eta) = e^{-i \eta X},
\end{equation}
where $X$ is the Hermitian generator of the transformation.
To quantify the change in $\rho$, we compute the fidelity between the initial and transformed states. For a pure initial state $\rho = \ket{\psi}\bra{\psi}$, the fidelity can be expressed as
\begin{equation}
    F_{\rho} (V(\eta)) \equiv \Tr \left[ \rho V(\eta) \rho V^{\dagger}(\eta) \right] \, .
\end{equation}
The optimal state(s) for detecting the transformation $V(\eta)$ undergone by the system are those that minimize $F_{\rho} (V(\eta))$. 

We define now the $\SUt$-average fidelity $\bar{F}_{[\rho]} (V)$ of a pure state $\rho$, transformed by a unitary matrix $V$, as
\begin{equation}
\label{PrhoQ}
\bar{F}_{[\rho]} (V)=\int_{\SUt} \diff \mu \,\Tr
\left(
\rho_U V \rho_U V^\dagger
\right)
\, ,
\end{equation}
where $\rho_U\equiv U \rho \, U^\dagger$, $U$ is a spin-$s$ SU(2) matrix, and $\td \mu=\td \mu(U)$ is the Haar measure of $\SUt$. Note that $\bar{F}_{[\rho]} (V)$ is invariant under the action of $\SUt$ (by conjugation) on either $\rho$ or $V$. Hence, it is a function of the $\SUt$ equivalence class of the state $\rho$, denoted by square brackets. We only show this explicitly for $\rho$, but it also holds for $V$, \ie, $\bar{F}_{[\rho]}(V)=\bar{F}_{\rho}([V])=\bar{F}_{[\rho]}([V])$. Thus, the $\SUt$-average fidelity admits two different interpretations: either the orientation of the probe system is averaged over its $\SUt$-orbit, or an average is performed over all transformations in the SU(2)-orbit of $V$.

Our fidelity criterion can be viewed as an SU(2)-averaged Loschmidt echo, where the unitary operator $V$ plays the role of the relative evolution between an unperturbed and a perturbed dynamics. In this language, optimal sensors are precisely those states that minimize the averaged echo, \ie, those whose dynamics is most sensitive to the applied perturbation.

For an infinitesimal parameter $\eta \ll 1$, one can calculate instead the quantum Fisher information (QFI), related to the first non-trivial term of the Taylor expansion of the fidelity~\cite{PhysRevA.111.022435,helstrom_quantum_1969,PhysRevLett.72.3439,zhou2019}
\begin{equation}
\label{Eq.QFI.dev}
    I_{\rho}(X) \equiv \left. -2  \frac{\partial^2 F_{\rho}(V(\eta))}{\partial \eta^2} \right|_{\eta=0} .
\end{equation}
Thus, $I_{\rho}(X)$ quantifies the sensitivity of $\rho$ to infinitesimal parameter changes associated with the generator $X$. \rema{For pure states, one recovers the well-known result that the QFI is four times the variance of the generator,}
\begin{equation}
\label{Eq.QFI}
    I_{\rho}(X) 
    = 
    4\big(\Tr(\rho X^2) -\Tr(\rho X)^2 \big) \equiv  \, 
    4 (\Delta_\rho X)^2 .
\end{equation}
\rema{Thus, in the infinitesimal regime, optimal sensors are those states that maximize the QFI with respect to the generator $X$.} Similarly, we define the $\SUt$-average QFI as
\begin{equation}
\label{Eq.Av.QFI}
	\bar{I}_{[\rho]}(X)
    =
    \int_{\SUt}\diff \mu \, 
    I_{U\rho U^{\dagger}}(X)\, ,
\end{equation}
where, again, $\bar{I}_{[\rho]}(X)=\bar{I}_{\rho}([X])$. Thus, we have the same dual interpretation as for finite transformations: $\bar{I}_{[\rho]}(X)$ can be thought of as averaged over states and/or generators, uniformly distributed (in the Haar sense) on their $\SUt$-orbit.

\subsection{\rema{Relevant classes of generators and applications}}
\label{SubSec.ApplicationsGenerators} 
\remal{We now specialize the general framework above to the classes of transformations that arise most often in applications. The first and most standard case is that of rotations generated by Hamiltonians linear in the angular-momentum operators,
\begin{equation}
H=\omega\,\hat{n}\cdot \mathbf{S},
\end{equation}
which we refer to as linear Hamiltonians. Here, $\mathbf{S}=(S_x,S_y,S_z)$, with $S_i$ the spin operators, and $\hat n$ specifies the rotation axis. A second important class, also ubiquitous in spin systems, is formed by quadratic Hamiltonians. These describe transformations whose generators are second order in the spin operators and can be written in the general form
\begin{equation}
\label{Eq.Quadratic.Ham}
H_Q 
= 
\mathbf{S} \,  Q \, \mathbf{S}^T\, ,
\end{equation}
where $Q$ is a real symmetric $3\times3$ matrix. As shown in Appendix~\ref{App.Quadrupolar}, up to a rotation, any traceless quadratic Hamiltonian can be parametrized by a linear combination of two generators,
\begin{equation}
\label{Eq.Gen.Quadratic.unipara}
H_1 = S_z^2 - \frac{s(s+1)}{3}\mathds{1}\, , \qquad
H_2 =  S_x^2 - S_y^2\,  .
\end{equation}
These two quadratic generators cover several physically relevant situations. For example, $H_1$ appears in quadrupolar couplings between an electric-field gradient and the spin of the probe, such as the coupling to the nuclear charge distribution~\cite{Asaad2020}. In this setting, the matrix $Q$ has entries $Q_{ij}\propto q_n V_{ij}$, where $V_{ij}=\partial^2 V/\partial x^i\partial x^j$ is the electric-field-gradient tensor and $q_n$ is the nuclear quadrupole moment. The same generator also appears as the one-axis twisting Hamiltonian, widely used to produce spin squeezing~\cite{MA201189}. This interaction is associated with a single distinguished axis.}

\remal{By contrast, $H_2$ corresponds to a two-axis twisting generator. Its orientation cannot be specified by a single axis alone, but requires a full orthogonal reference frame. This type of quadratic interaction is another standard mechanism for generating spin squeezing~\cite{MA201189}. When no ambiguity arises, we will simply refer to this interaction Hamiltonian as a squeezing generator.}

\rema{In many physical situations the generator contains both a linear Zeeman term and a quadratic anisotropy, $H=\omega\,\hat n\cdot\mathbf S+\mathbf S Q\mathbf S^T$. This mixed case describes, for instance, a spin probe subject simultaneously to a magnetic field and to quadrupolar or anisotropic interactions.}
\section{Preliminary concepts}
\label{MathPre}
\subsection{Notation and geometric background}
\label{Ban}
We denote the real vector space of $n$-dimensional Hermitian matrices by $\un \equiv \mathfrak{u}$, and endow it with the metric 
\begin{equation}
    \label{metricun}
    g(X,Y)
    =
    \frac{1}{2} \Tr(X Y)\, .
\end{equation}
The set (locus) $\mathbb{P}$ \rema{of the density matrices of pure states} in $\un$ lies on a sphere of radius $1/\sqrt{2}$, centered at the origin. In what follows, it will often prove convenient to vectorize a (not necessarily Hermitian) $r \times r$  matrix, $A \rightarrow \ket{A}$, where the $r^2$ entries of $A$ are arranged sequentially into a vector, reading them left-to-right and top-to-bottom. For example, $\ket{A}=(a,b,c,d)^T$ for $A = \left( \begin{smallmatrix} a & b \\ c & d \end{smallmatrix} \right)$, \ie, $\ket{A}_{(ij)}=A_{ij}$, where $(ij)$ is a composite index ranging over the $r^2$ pairs $\{11,12,\ldots,1r,21,\ldots,rr\}$ (we omit the parentheses from now on). Then, for $r \times r$ matrices $U_1$, $A$, $U_2$, we have
\begin{equation}
    \label{vecprod}
    \ket{U_1AU_2^\dagger}=(U_1 \otimes \bar{U}_2) \ket{A}
    \, ,
\end{equation}
where $U_2^\dagger=\bar{U}_2^T$, $\bar{U}_2$ is the complex conjugate of $U_2$, and $(A \otimes B)_{ij \, kl}=A_{ik} B_{jl}$. Note also that $\Tr(A B)=\braket{A^\dagger | B}$.  
We will in fact need to also vectorize $n^2 \times n^2$ matrices, for example, for $R \equiv \rho \otimes \bar{\rho}$ we have
\begin{equation*}
    \ket{R}_{ijkl}
    =
    R_{ij \, kl}
    =
    \rho_{ik} \bar{\rho}_{jl}
    \, .
\end{equation*}
Note that in the \lhs{} above, $ijkl$ is a composite 4-index, taking on the $n^4$ values $\{1111,\ldots,nnnn\}$, while in the middle $ij$ and $kl$ are both composite 2-indices, enumerating the $n^2$ rows and columns, respectively, of $R$.

\subsection{The subspace of \texorpdfstring{$\text{SU}(2)$}{Lg}-invariants}
\label{TsoSU2io}
We now proceed to construct $\text{SU}(2)$-invariants of operators, focusing first on $\rho$. The adjoint action of $\SUt$ on $\un$, $g \times u \mapsto U_g u \, U_g^\dagger$, where $U_g=D^{(s)}(g)$, may be vectorized as $\ket{u} \mapsto \mathbb{U}_g \ket{u}$, where $\mathbb{U}_g=U_g \otimes \bar{U}_g$. This action provides a representation of $\SUt$, which, as mentioned above, decomposes into a direct sum of irreducible representations with spin $l$ ranging from 0 to $2s$. The subspace $I^{\SUt}$ of matrices in $\un$ invariant under this action is 1-dimensional, corresponding to spin $l=0$, consisting of multiples of the identity matrix. One way of understanding this is via the familiar spin ``addition'' recipe, 
\begin{equation}
s_1 \otimes s_2 = (s_1 + s_2)  \oplus
(s_1 + s_2-1)  \oplus
\ldots \oplus |s_1-s_2| \, .    
\end{equation}
The space of $n$-dimensional complex matrices $M_n(\mathbb{C})$ may be seen as $\mathbb{C}^n \otimes \mathbb{C}^{n \, *}$ ($\mathbb{C}^{n \, *}$ being the dual of $\mathbb{C}^n$), since every such matrix $M$ can be written as $M = \sum M_{ij} \ket{e_i}\bra{e_j}$, where $\{ \ket{e_i}: i=1,\ldots,n\}$ is an orthonormal basis in $\mathbb{C}^n$. Since both $\mathbb{C}^n$ and $\mathbb{C}^{n \, *}$ transform in a spin-$s$ irreducible representation of $\SUt$, $M_n(\mathbb{C})$ decomposes according to 
\begin{equation}
    \label{spseq}
    s \otimes s= 2s \oplus \ldots \oplus 0
    \, .
\end{equation} 
Note that the only way to produce spin-0 components is by ``adding'' equal spins.  

The appearance of the adjoint action mentioned above is natural, since the transformation law $\ket{\psi'}=U \ket{\psi}$ leads to $\ket{\rho'}=\mathbb{U}\ket{\rho}$. In what follows we deal with the ``big density matrix'' $R=\ket{\rho}\bra{\rho}$ which transforms like 
\begin{equation}
    \label{Rtrans}
    R'=\mathbb{U} R \mathbb{U}^\dagger
    \, ,
    \quad
    \ket{R'}=(\mathbb{U} \otimes \bar{\mathbb{U}}) \ket{R}
    \, .
    \end{equation}
To connect with the language of superoperators, we note that $R$ is just the superoperator sending $A$ to $\Tr(\rho A)\rho$, acting on the vectorized argument $\ket{A}$.
For reasons that will soon become clear, we are interested in the subspace $\mathbb{I}^{\, \SUt}$ of $\mathfrak{u}(n^2)$ spanned by the vectors (\ie, $n^2$-dimensional Hermitian matrices) that are invariant under the above action. Reasoning as in the previous case, we inquire about how many spin-0's are produced in the addition of four spin-$s$'s. According to a remark made above, this number is equal to the number of spin-$s$'s that appear in the sum of three $s$'s. From~(\ref{spseq}) we see that adding $s$ to each of the terms in the \rhs{} produces exactly one $s$ --- we conclude that $\dim \mathbb{I}^{\, \SUt}=2s+1$, since it can be shown that all resulting invariants are linearly independent. To find a basis of $\mathbb{I}^{\, \SUt}$, we note that if the $T_{lm}$'s transform as spinors (see appendix~\ref{Sub.Polarization} for the definition of the matrices $T_{lm}$ and, in particular, Eq.~\eqref{Tlmirrep} for their transformation properties), and, hence, the $\bar{T}_{lm}$ as conjugate spinors, then the operators 
\begin{equation}
    \label{TTlm}
    \mathbb{T}_{l}=\sum_{m=-l}^l T_{lm} \otimes \bar{T}_{lm}
    \, ,
    \quad l=0,\ldots,2s
    \, ,
\end{equation}
are invariant, for the same reason $\braket{\psi|\psi}$ is,
and so they provide a basis in $\mathbb{I}^{\, \SUt}$, as they can be shown to be an orthogonal set --- they are also Hermitian. Note that the index $l$ in $\mathbb{T}_l$ does not refer to its $\SUt$ transformation properties (all $\mathbb{T}_l$'s should carry an index 0 in that respect), but, rather, it is a reminder of the $T_{lm}$'s used in~\eqref{TTlm}. The operator $\mathbb{T}_{l}$ is the superoperator sending $A$ to $\sum_{m=-l}^l T_{lm} \,A \, T_{lm}^\dagger$, acting on the vectorized argument $\ket{A}$. It is also easy to see that the $\mathbb{T}_l$'s commute among themselves. Indeed, working in the $T$-basis, note that multiplication of the spinor $\ket{T_{l}}$ (with components the operators $T_{lm}$)  by the spin-0 operator $(\mathbb{T}_l)_T$ (\ie, $\mathbb{T}_l$ in the $T$-basis) must transform as $ 0 \otimes l=l$. Since there is only one spin-$l$ multiplet, it must hold that 
\begin{equation}
    \label{TTLTlp}
    (\mathbb{T}_{l})_T \ket{T_{l'm'}}_T=\lambda_{ll'} \ket{T_{l'm'}}_T
    \, ,
\end{equation}
for some constants $\lambda_{ll'}$, a detailed description of which is given in appendix \ref{App.prop:lambda}. This means that all $(\mathbb{T}_l)_T$ are simultaneously diagonal, and proportional to the identity matrix in each $l$-subspace (as implied by Schur's lemma),
\begin{equation}
    \label{TbbTbasis}
    (\mathbb{T}_l)_T=\bigoplus_{l'=0}^{2s} \lambda_{ll'} I_{2l'+1}
    \, ,
\end{equation}
and hence they commute among themselves in every basis.

An alternative basis in $\mathbb{I}^{\, \SUt}$ is $\{\mathbf{T}_l:l=0,\ldots,2s\}$, where
\begin{equation}
    \label{Tbfdef}
    \mathbf{T}_l=\sum_{m=-l}^l \ket{T_{lm}}\bra{T_{lm}}
    \, ,
\end{equation}
the invariance of which follows from an argument analogous to the one used above for the $\mathbb{T}_l$.
The matrix $(\mathbf{T}_l)_T$ (\ie, $\mathbf{T}_l$ in the $T$-basis) is (inevitably) diagonal, with a $(2l+1)$-dimensional identity matrix in the $l$-subspace, and zeros elsewhere. The $\mathbf{T}_l$'s are projection operators onto the irreducible subspaces, satisfying
\begin{equation}
    \label{Tbfprop}
    \mathbf{T}_l \mathbf{T}_{l'}=\delta_{ll'} \mathbf{T}_l
    \, , \quad
    \sum_{l=0}^{2s} \mathbf{T}_l=I_{n^2}
    \, ,
\end{equation}
relations that remain valid in all bases.
Note that
\begin{equation}
    \label{rhoTrho}
    \bra{\rho} \mathbf{T}_l \ket{\rho}=\sum_{m=-l}^l |\rho_{lm}|^2 \equiv |\rho_l|^2
    \, .
\end{equation}
The relation between the two bases is given by 
\begin{equation}
    \label{TbfTbb}
    \mathbb{T}_l=\sum_{l'=0}^{2s} \lambda_{ll'} \mathbf{T}_{l'}
    \, ,
    \quad
    \mathbf{T}_l=\sum_{l'=0}^{2s} \lambda_{ll'} \mathbb{T}_{l'}
    \, ,
\end{equation}
the first of which  is obvious from~(\ref{TbbTbasis}), while the second follows from the property of the $\lambda$-matrix $\lambda^2=I$ (see appendix~\ref{App.prop:lambda}). Both bases are orthogonal,
\begin{equation}
    \label{TbbTbfortho}
    \braket{\mathbb{T}_l|\mathbb{T}_{l'}}
    =(2l+1) \delta_{ll'}=
    \braket{\mathbf{T}_l|\mathbf{T}_{l'}}
    \, . 
\end{equation}
It will prove convenient in the following to work with the bases $\{\tilde{\mathbb{T}}_l \equiv \mathbb{T}_{\tilde{l}}=\mathbb{T}_l/\sqrt{2l+1} \}$ and $\{\tilde{\mathbf{T}}_l \equiv \mathbf{T}_{\tilde{l}}=\mathbf{T}_l/\sqrt{2l+1} \}$.
The projector $\Lambda$ from $\mathfrak{u}(n^2)$ to $\mathbb{I}^{\, \SUt}$ is given by
\begin{equation}
    \label{ISU2proj}
    \Lambda
    =
    \sum_{l=0}^{2s} \ket{\mathbb{T}_{\tilde{l}}} \bra{\mathbb{T}_{\tilde{l}}}
    =
    \sum_{l=0}^{2s} \ket{\mathbf{T}_{\tilde{l}}} \bra{\mathbf{T}_{\tilde{l}}}
    \, .
\end{equation}

\remal{We now define the \emph{$l$-coherence} of $\rho$~\footnote{It is important to note that the notion of coherence used here differs from that considered in the resource theory of coherence~\cite{Streltsov2017}, where measures of coherence (such as the $\ell_1$ norm) quantify the amount of quantum superposition of a mixed state with respect to a fixed reference basis.}, $l=0,\ldots,2s$, via 
\begin{equation}
    \label{rldef}
    r_l(\rho)
    \equiv
  \sum_{m=-l}^l  \Tr(\rho T_{lm} \rho T_{lm}^\dagger)
  =
  \bra{\rho} \mathbb{T}_l \ket{\rho}
  \, ,
\end{equation}
and collect all $r_{l}$'s in an $n$-dimensional vector $r_{\rho}=(r_0,\ldots,r_{2s})$ of $\SUt$-invariant functions on $\mathbb{P}$ --- note that we write just $r_l$ in what follows, dropping the explicit mention of $\rho$.
Using the first equality of~(\ref{TbfTbb}) and~(\ref{rhoTrho}) we infer that $r_l=\sum_{l'=0}^{2s}\lambda_{ll'} |\rho_{l'}|^2$. On the other hand, from~(\ref{rldef}) we get
\begin{align}
\label{Eq.rl.to.rhol}
    r_l
    =
    \sum_{m=-l}^l \bra{\psi} T_{lm} \ket{\psi} \bra{\psi} T_{lm}^\dagger \ket{\psi}
=
    \sum_{m=-l}^l \bar{\rho}_{lm} \rho_{lm}
    =
    |\rho_l|^2
    \, ,
\end{align}
implying 
\begin{equation}
    \label{reigen}
    \lambda r_{\rho} =r_{\rho}
    \, ,
\end{equation}
(where $\lambda$ is a matrix and $r$ a vector) and
\begin{equation}
    \label{rhoTTlrho}
    \bra{\rho} \mathbb{T}_l \ket{\rho}=
    (r_{\rho})_l
    =
    \bra{\rho} \mathbf{T}_l \ket{\rho}
    \, .
\end{equation}
From now on, we write our expressions in terms of the $\SUt$-invariants \rema{$(r_{\rho})_l$, or just $r_l$ for short}, and extend the definition of an SU(2)-invariant $r$-vector to general operators $M$ via
\begin{equation}
    (r_{M})_{l}
    =
    \langle M|\mathbf{T}_{l}|M\rangle
    \, ,
    \quad l=0,\ldots,2s
    \, .
    \label{rXvec}
\end{equation}
}
%
%
%
\subsection{Further concepts related to spin states}
We conclude this section of preliminary concepts with a brief introduction to anticoherence and the Majorana representation of spin states. Comprehensive discussions of these topics can be found in, e.g., \cite{Zim:06,PhysRevLett.114.080401,PhysRevA.92.052333,Bou.etal:17,PhysRevA.96.032304} for anticoherence, and \cite{Maj:32,Chr.Guz.Ser:18,Aulbach_2010,PhysRevA.81.062347,Björk_2015,PhysRevA.85.032314,Bou.etal:17} for the Majorana representation. 
A pure state $\rho = \ket{\psi}\bra{\psi}$ is said to be \emph{$t$-anticoherent}~\cite{Zim:06,PhysRevLett.114.080401} if $\rho_{lm} = 0$ for $l = 1, \dots, t$ and $m = -l, \dots, l$. \rema{If} $\rho$ is $t$-anticoherent, every state in its $\SUt$ orbit is also $t$-anticoherent. 

The Majorana representation (also called \emph{stellar} representation) for pure spin-$s$ states~\cite{Maj:32,Chr.Guz.Ser:18} provides a bijective mapping between physical states $[\psi] \in \mathds{C}P^{2s}$ to $N = 2s$ unordered points on the sphere $S^2$, known as the \emph{constellation} of $[\psi]$. In this representation, rotations (\ie, $\text{SU}(2)$ transformations) of the state $[\psi]$ correspond directly to physical rotations of its constellation. 
\begin{figure}[t]
\centering
\tikzset{every picture/.style={line width=0.75pt}} 

\begin{tikzpicture}[x=0.75pt,y=0.75pt,yscale=-1,xscale=1]

\draw   (373.09,145.54) -- (380.67,143.16) -- (383.05,150.75) -- (375.46,153.13) -- cycle ;
\draw   (249.79,184.38) -- (257.38,182) -- (259.76,189.59) -- (252.17,191.96) -- cycle ;
\draw [color={rgb, 255:red, 0; green, 65; blue, 255 }  ,draw opacity=1 ][line width=0.75]    (174.5,216.46) -- (403.19,144.42) ;
\draw [shift={(406.05,143.52)}, rotate = 162.51] [fill={rgb, 255:red, 0; green, 65; blue, 255 }  ,fill opacity=1 ][line width=0.08]  [draw opacity=0] (8.93,-4.29) -- (0,0) -- (8.93,4.29) -- cycle    ;
\draw    (381.63,54.92) .. controls (346.85,108.32) and (356.45,124.85) .. (402.4,114.2) ;
\draw  [color={rgb, 255:red, 65; green, 117; blue, 5 }  ,draw opacity=1 ][fill={rgb, 255:red, 0; green, 0; blue, 0 }  ,fill opacity=1 ] (379.01,54.92) .. controls (379.01,53.47) and (380.19,52.3) .. (381.63,52.3) .. controls (383.08,52.3) and (384.25,53.47) .. (384.25,54.92) .. controls (384.25,56.36) and (383.08,57.53) .. (381.63,57.53) .. controls (380.19,57.53) and (379.01,56.36) .. (379.01,54.92) -- cycle ;
\draw  [color={rgb, 255:red, 65; green, 117; blue, 5 }  ,draw opacity=1 ][fill={rgb, 255:red, 0; green, 0; blue, 0 }  ,fill opacity=1 ] (399.78,114.2) .. controls (399.78,112.76) and (400.95,111.58) .. (402.4,111.58) .. controls (403.85,111.58) and (405.02,112.76) .. (405.02,114.2) .. controls (405.02,115.64) and (403.85,116.82) .. (402.4,116.82) .. controls (400.95,116.82) and (399.78,115.64) .. (399.78,114.2) -- cycle ;
\draw  [color={rgb, 255:red, 65; green, 117; blue, 5 }  ,draw opacity=1 ][fill={rgb, 255:red, 65; green, 117; blue, 5 }  ,fill opacity=1 ] (358.86,106.94) .. controls (358.86,105.5) and (360.04,104.32) .. (361.48,104.32) .. controls (362.93,104.32) and (364.1,105.5) .. (364.1,106.94) .. controls (364.1,108.39) and (362.93,109.56) .. (361.48,109.56) .. controls (360.04,109.56) and (358.86,108.39) .. (358.86,106.94) -- cycle ;
\draw  [color={rgb, 255:red, 0; green, 0; blue, 0 }  ,draw opacity=1 ][fill={rgb, 255:red, 155; green, 155; blue, 155 }  ,fill opacity=0.35 ][line width=0.75]  (294.67,80.32) .. controls (299.59,77.39) and (312.72,65.99) .. (324.84,88.8) .. controls (336.95,111.61) and (312.68,115.5) .. (295.33,137.5) .. controls (277.97,159.49) and (253.62,161.56) .. (237.25,137.03) .. controls (220.87,112.51) and (250.64,103.41) .. (269.08,95.09) .. controls (287.53,86.77) and (289.76,83.25) .. (294.67,80.32) -- cycle ;
\draw  [color={rgb, 255:red, 65; green, 117; blue, 5 }  ,draw opacity=1 ][fill={rgb, 255:red, 65; green, 117; blue, 5 }  ,fill opacity=1 ] (230.01,128.23) .. controls (230.01,126.79) and (231.18,125.62) .. (232.63,125.62) .. controls (234.07,125.62) and (235.25,126.79) .. (235.25,128.23) .. controls (235.25,129.68) and (234.07,130.85) .. (232.63,130.85) .. controls (231.18,130.85) and (230.01,129.68) .. (230.01,128.23) -- cycle ;
\draw    (174.5,216.46) -- (174.06,35.45) ;
\draw [shift={(174.05,32.45)}, rotate = 89.86] [fill={rgb, 255:red, 0; green, 0; blue, 0 }  ][line width=0.08]  [draw opacity=0] (8.93,-4.29) -- (0,0) -- (8.93,4.29) -- cycle    ;
\draw    (174.5,216.46) -- (411.05,216.45) ;
\draw [shift={(414.05,216.45)}, rotate = 180] [fill={rgb, 255:red, 0; green, 0; blue, 0 }  ][line width=0.08]  [draw opacity=0] (8.93,-4.29) -- (0,0) -- (8.93,4.29) -- cycle    ;
\draw [color={rgb, 255:red, 65; green, 117; blue, 5 }  ,draw opacity=1 ] [dash pattern={on 4.5pt off 4.5pt}]  (256.48,205.63) -- (216.48,78.03) ;
\draw [color={rgb, 255:red, 65; green, 117; blue, 5 }  ,draw opacity=1 ] [dash pattern={on 4.5pt off 4.5pt}]  (379.28,165.23) -- (339.28,37.63) ;
\draw [color={rgb, 255:red, 245; green, 140; blue, 35 }  ,draw opacity=1 ]   (174.5,216.46) -- (228.68,132.94) ;
\draw [shift={(230.31,130.43)}, rotate = 122.97] [fill={rgb, 255:red, 245; green, 140; blue, 35 }  ,fill opacity=1 ][line width=0.08]  [draw opacity=0] (8.93,-4.29) -- (0,0) -- (8.93,4.29) -- cycle    ;

\draw (280.52,92.43) node [anchor=north west][inner sep=0.75pt]  [font=\large]  {$\mathcal{R}^{( s)}$};
\draw (409.6,139.32) node [anchor=north west][inner sep=0.75pt]    {$\tilde{r}_{V}$};
\draw (418,210) node [anchor=north west][inner sep=0.75pt]    {$\tilde{r}_{1}$};
\draw (169.2,15.2) node [anchor=north west][inner sep=0.75pt]    {$\tilde{r}_{2}$};
\draw (227,165.8) node [anchor=north west][inner sep=0.75pt]    {$\Pi _{1}$};
\draw (350.62,127.4) node [anchor=north west][inner sep=0.75pt]    {$\Pi _{2}$};
\draw (189.91,150.19) node [anchor=north west][inner sep=0.75pt]    {$\tilde{r}_{\rho }$};
\end{tikzpicture}
\caption{Sketch illustrating our geometric approach to calculating the optimal SU(2)-averaged sensors of \rema{transformations $V$}. The SU(2)-average fidelity of the pure state $\rho$, being transformed by $V$, is given by the inner product of the invariant vectors $\tilde{r}_\rho$ (orange arrow) and $\tilde{r}_V$ (blue arrow). For a given $V$ (and, hence, a fixed $\tilde{r}_V$), we look for a state $\rho$ (which ranges over the locus $\mathcal{R}^{(s)}$, in gray) that minimizes $\tilde{r}_\rho \cdot \tilde{r}_V$. As schematically shown, the problem can be formulated by considering planes $\Pi$ of constant height with respect to the direction of $\tilde{r}_{V}$ (orthogonal complements to $\tilde{r}_V$). Because $\mathcal{R}^{(s)}$ has maximal dimension in the $\rema{r_0=1/n}$ hyperplane inside the positive eigenspace of the matrix $\tilde{\lambda}$ (this is the plane sketched in the figure), its lowest point, in the $\tilde{r}_V$ direction, \rema{always lies} on its boundary (where $\Pi_1$ is tangent to $\mathcal{R}^{(s)}$ in the figure). We also sketch a hypothetical locus of lower dimension (curve on the right), the lowest point of which lies in its interior (with respect to the intrinsic topology).}
\label{Fig.1.sketch}
\end{figure}
\section{Optimal sensors for arbitrary transformations}
\label{Sec.Ofbs}
\subsection{\remal{A general formula for the average fidelity}}
In this section we obtain a particularly simple general formula for the $\SUt$-average fidelity corresponding to a finite unitary transformation $V$, as defined in~\eqref{PrhoQ}. By successive vectorizations we compute
\begin{align}
    \bar{F}_{[\rho]} (V)
    &=
    \int_{\SUt} \diff \mu \bra{\rho_U} V \otimes \bar{V} \ket{\rho_U}
    \nonumber
    \\
    &=
    \int_{\SUt} \diff \mu \bra{\rho} \mathbb{U}^\dagger ( V \otimes \bar{ V}) \mathbb{U} \ket{\rho}
    \nonumber
    \\
    &=
    \int_{\SUt} \diff \mu \,\Tr \left( 
    ( V \otimes \bar{ V})\mathbb{U} R\mathbb{U}^\dagger
    \right)
    \nonumber
    \\
    &=
    \int_{\SUt} \diff \mu
    \bra{ V^\dagger \otimes  V^T} \mathbb{U} \otimes \bar{\mathbb{U}} \ket{R}
    \nonumber
    \\
    &=\bra{V^{\dagger}\otimes V^{T}}\left(\int_{\SUt}\diff \mu \mathbb{U}\otimes \bar{\mathbb{U}}\right) \ket{R}\nonumber\, ,
    \end{align}
where $R\equiv \ket{\rho} \bra{\rho}$ and $[\rho]$ denotes the $\text{SU}(2)$ orbit of $\rho$. The integral that appears above is known to be given by 
\begin{equation}
    \label{bbUbbUint}
    \int_{\SUt} \diff \mu \mathbb{U} \otimes \bar{\mathbb{U}}=\Lambda
    \, ,
\end{equation}
with $\Lambda$ as in~(\ref{ISU2proj}) (see, \eg,~\cite{Ful.Har:04}). In this way the average fidelity in~(\ref{PrhoQ}) reduces to that of a simple Euclidean inner product between two vectors, the components of which are SU(2) invariants of $V$ and $\rho$:
    \begin{align}
    \bar{F}_{[\rho]} (V)
    &=
    \bra{V^{\dagger}\otimes V^{T}}\Lambda\ket{R}
    \nonumber 
    \\
    &= 
    \sum_{l=0}^{2s} \frac{(r_{V})_{l}(r_{\rho})_{l}}{2l+1}
    \nonumber
    \\
    &=
    \tilde{r}_{V} \cdot \tilde{r}_\rho
    \, ,
    \label{PrQrrho}
\end{align}
where 
\begin{equation}
\label{Eq.Invariant.vector}
\tilde{r}_\rho \equiv(\tilde{r}_{0},\ldots, \tilde{r}_{2s})\quad \text{with} \quad
\tilde{r}_{l}\equiv \frac{r_l}{\sqrt{2l+1}}
\, ,    
\end{equation}
and similarly for $\tilde{r}_{V}$.

Expanding~(\ref{PrQrrho}) to second order in the transformation parameter $\eta$ (with $V=e^{-i\eta X}$), and using~(\ref{Eq.Av.QFI}), we obtain the average QFI:
\begin{equation}
    \label{PrhoZt}
    \bar{I}_{[\rho]}(X)
    = 4 \left( \frac{\Tr(X^2)}{2s+1} - \tilde{r}_\rho \cdot \tilde{r}_{X}
    \right)  \, .
\end{equation}

For a given unitary operator $V$ the optimal sensor states are those minimizing the inner product (\ref{PrQrrho}) --- similarly, for a given Hermitian generator $X$, they are those minimizing the inner product appearing in~(\ref{PrhoZt}). 
\subsection{\remal{The locus of \texorpdfstring{$r_\rho$}{Lg}} \rema{and universality of optimal sensors}}
The vector space where the invariant vectors $r_\rho$ live (with $\rho$ an $n$-dimensional pure density matrix, $n=2s+1$), is $\mathbb{R}^n$. Eq.~(\ref{reigen}) imposes $\text{rank}(\lambda-I)=\lceil s \rceil$ independent linear constraints on $r_\rho$, and, additionally, $r_0$ is constant, which implies that the dimension of the locus $\mathcal{R}^{(s)}$ is at most $\lfloor s \rfloor$. In Appendix~\ref{App.dimproof} it is shown that this upper bound is actually saturated for all $s$, so that $\text{dim} \, \mathcal{R}^{(s)}=\lfloor s \rfloor$, \emph{i.e.}, $\mathcal{R}^{(s)}$ is of maximal dimension in the hyperplane (inside the positive eigenspace of $\lambda$) defined by $\rema{r_0=1/n}$. The basic idea for the proof is the identification of a parametrized family of states that projects to a region of $\mathcal{R}^{(s)}$ with dimension $\lfloor s \rfloor$, as sketched in Fig.~\ref{Fig.1.sketch}.

Given that $\mathcal{R}^{(s)}$ lies in the $\rema{r_0=1/n}$ hyperplane inside the positive eigenspace of $\lambda$, and is of maximal dimension there, the \rema{r-vectors} corresponding to optimal sensors lie on the boundary of $\mathcal{R}^{(s)}$ (for a sketch see Fig.~\ref{Fig.1.sketch}) --- the only exception being the case in which $\tilde{r}_V$ is orthogonal to $\mathcal{R}^{(s)}$
 (see, e.g., the $r$-vectors corresponding to squeezing in Fig.~\ref{Fig.4.spin32locus}). States with $r_\rho$ in the interior of $\mathcal{R}^{(s)}$ will never be optimal sensors, no matter what $V$ is (with the exception mentioned above). If $r_{V}$ is taken to define the ``up'' direction in $r$-space, the optimal sensors are the lowest-lying states, and the worst sensors are the highest. Note that these statements are valid for any value of $s$, regardless of the specific form of $\mathcal{R}^{(s)}$. 
\rema{We are thus led  to the following}

\noindent
\rema{\paragraph*{ \textbf{Main result.}} \emph{For almost all unitary transformations $V$, the $r$-vectors of optimal (and worst) sensor states lie on the boundary of $\mathcal{R}^{(s)}$. Thus, optimal and worst sensors, for almost any unitary transformation, form a zero-measure subset of quantum states.}}

When $\mathcal{R}^{(s)}$ is a polytope, as it turns out to be for spins up to 2, the optimal (as well as the worst) sensors are generically to be found among the vertices of the polytope, and this fact is independent of the nature of $V$. For the special class of operators $V$, such that $\tilde{r}_V$ becomes orthogonal to a face of  $\mathcal{R}^{(s)}$, that entire face may correspond to optimal sensors. When that happens, small variations of $V$ around those special values result in discontinuous changes of the corresponding optimal sensors. This behavior has been observed before~\cite{PhysRevA.95.052125,Mar.Wei.Gir:20}, and was, until now, quite puzzling, but becomes obvious once the problem is cast in geometric terms as above. 
\section{Optimal fidelity-based sensors}
\label{Sec.Gen.Uni}
\remal{
We now apply our approach (summarized in Eq.~\eqref{PrQrrho}) to calculate the optimal sensor states \rema{for several unitary transformations}, for arbitrary spin in the infinitesimal case and for $s=1,\,3/2,\,2,\,5/2$ in the finite case.}


\subsection{\remal{Infinitesimal case: Average QFI}}
\remal{In the special case where $X=\Gamma_{\ell m}$ (see~(\ref{Hbndef}) for the definition of the generators $\Gamma_{\ell m}$), we have $(r_{X})_{k}=2\delta_{\ell k}$, so that
\begin{equation}
\bar{I}_{[\rho]}(\Gamma_{\ell m})
=8\left( \frac{1}{2s+1} - \frac{r_{\ell}}{2\ell+1}
\right) 
\, .
\label{eq:I_lm}
\end{equation}
Therefore, the optimal sensors of an infinitesimal transformation generated by $ \Gamma_{\ell m}$ are the states \rema{minimizing $r_{\ell}$. When the minimum value zero is attainable, these are the states with $r_{\ell}=0$. In particular, for rotations ($\ell=1$) and $s\geq 1$, the optimal sensors are the 1-anticoherent states}. These results, associated with rotations, are in accordance with those previously derived in other references~\cite{PhysRevA.95.052125,Mar.Wei.Gir:20,Bou.etal:17}.}

\remal{Now, consider the general quadratic traceless Hamiltonian~\eqref{Eq.Quadratic.Ham}. Using appropriate rotations and normalization (see Appendix~\ref{App.Quadrupolar}), we arrive at the general form
\begin{equation}
\label{Eq.Ten.Quadratic}
    H(\alpha) = 
    \cos \alpha \, \Gamma_{20} + \sin \alpha \, \Gamma_{22} 
    \, ,
\end{equation}
with $\alpha \in [0,\pi/2]$. Note that $H(0) \propto S_z^2 - s(s+1)\mathds{1}/3$ and $H(\pi/2)\propto \left(S_x^2 - S_y^2 \right)$. A short calculation shows  that the average \rema{QFI}~\eqref{PrhoZt} is independent of $\alpha$, 
\begin{equation}
	\label{eq:PrhoZt_Halpha}
	\bar{I}_{[\rho]}\bigl(H(\alpha)\bigr)
	=
	8\left(\frac{1}{2s+1}-\frac{r_2}{5}\right)
    \, ,
\end{equation}
as was expected from the result in Eq.~\eqref{eq:I_lm}.}
\begin{figure}[t!]
	\centering
	\includegraphics[scale=0.7]
    {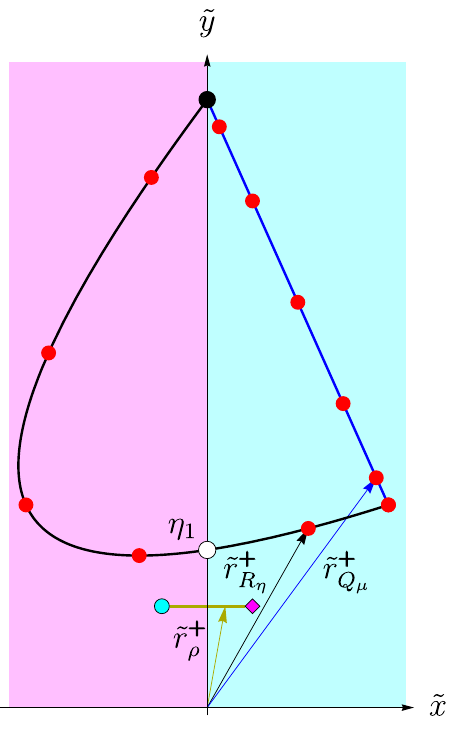}
	\caption{
    Locus $\mathcal{R}^{(1)}$ (horizontal mustard yellow segment) and curves corresponding to rotations $R_\eta$ (black) and squeezing $Q_\mu$ (blue) in the $(\tilde{x},\tilde{y})$-plane. The cyan disc on the left end of $\mathcal{R}^{(1)}$ stands for the coherent state while the magenta square on the right end denotes the anticoherent one. For any transformation $V$ such that the $\phi$ coordinate of $\tilde{r}^{+}_{V}$ is less than $\pi/2$, the optimal $V$-sensor is the coherent state. \rema{While} for $\phi>\pi/2$, \rema{the anticoherent state is optimal}. The little white disc on the rotation curve corresponds to the critical rotation angle $\eta_1=\arccos(-2/3)$ for which the optimal sensor changes discontinuously. Little red dots on both curves correspond to $30^\circ$ parameter increments. The black dot at the top corresponds to the identity transformation. Note that both curves are traversed twice, from the identity at the top to the red dot at the ``corner'' on the right for $0 \leq \eta,\mu \leq \pi$, and  retracing the same locus back to the identity for $\pi \leq \eta,\mu \leq 2\pi$. Special to the $s=1$ case, the preimage of any point of $\mathcal{R}^{(1)}$ in projective space is a single SU$(2)$ orbit, \ie, only states with Majorana constellations of the same shape project to the same point of $\mathcal{R}^{(1)}$.}
\label{Fig.3.spin1locus}
\end{figure}
\subsection{Finite case}
\label{SubSec.Finite}

A further refinement of~(\ref{PrQrrho}) is possible, based on~(\ref{reigen}). As mentioned above, and proved in Appendix~\ref{App.prop:lambda}, the matrix $\lambda$ that was introduced in~(\ref{TTLTlp}) satisfies the identity $\lambda^2=I$, which means its eigenvalues are $\pm 1$, and $r_\rho$ lives in the positive eigenspace. Given that $\Tr \, \lambda=2s+1 \, (\text{mod} \, 2)$ (see Appendix~\ref{App.prop:lambda}), there are as many positive eigenvalues of $\lambda$ as there are negative, when $s$ is half-integer, and one more positive than negative, when $s$ is integer. Accordingly, by switching to the eigenbasis of $\lambda$ we work with (roughly) half the components of $r_\rho$, those of its positive part, the rest being zero --- this simplifies the analysis considerably, and allows the visualization of $\mathcal{R}^{(s)}$ for relatively high spin values (we go up to $s=5/2$ in this article, but $\mathcal{R}^{(s)}$ can, in principle, be visualized for $s$ up to 7/2 --- we explain how below). 

It will prove convenient to implement the above considerations \rema{by} working in the $\tilde{\mathbb{T}}$-basis of $\mathbb{I}^{\, \SUt}$ that appears in~(\ref{ISU2proj}), because in that basis $\lambda$ becomes symmetric.
Indeed, repeating the calculation that led to~(\ref{PrQrrho}) with a general Hermitian matrix $B$ replacing $\rho = \ket{\psi} \bra{\psi}$, we find
\begin{equation}
    \bar{F}_{[B]}(V) = G(r_B,r_{V})
    \, ,
    \label{PBQ}
\end{equation}
where $(G_{l l'})$ is a natural metric in $r$-space, with $G_{l l'}=\lambda_{ll'}/(2l+1)$ --- that $G$ is a symmetric matrix follows from~(\ref{eq:prop_lambda_1}). This, more general, result reduces to~(\ref{PrQrrho}) when $B$ is a rank-1 projector. Note also that $\bar{F}_{[B]}(V) = \bar{F}_{[V]}(B) $, and that $G=A^2 \lambda$, where $A$ is diagonal, with $A_{\ell \ell}=1/\sqrt{2\ell+1}$. Switching to the \rema{basis} 
$\{ \mathbb{T}_{\tilde{l}} \}$ the metric transforms to $\tilde{G}=A^{-1} G A^{-1}=A \lambda A^{-1}$, \ie, $\tilde{G}$ is a matrix similar to $\lambda$. Under the same change of basis, $\lambda$ transforms as $\tilde{\lambda}=A \lambda A^{-1}=\tilde{G}$ --- in other bases, the two matrices are, in general, different %
\footnote{%
The reason $G$, $\lambda$ transform differently under a change of basis is that the former maps pairs of vectors to numbers, while the latter maps vectors to vectors.%
}. 
In the $\tilde{\lambda}$-eigenbasis, the metric becomes  $\tilde{G}^D=\text{diag}(-1,\ldots,-1,1,\ldots,1)$.

When $\rho$ is a pure state, $\tilde{r}_\rho=A r_\rho$ lives in the positive eigenspace of $\tilde{\lambda}$. When referred to the $\tilde{\lambda}$-eigenbasis,  it takes  the form 
\begin{equation}
\label{Eq.Vector.zeronegative}
\tilde{r}_\rho^D=(0,\ldots,0,\tilde{r}^{+}_1,\ldots,\tilde{r}^{+}_{\lfloor s\rfloor +1})\equiv (\mathbf{0},\tilde{r}_\rho^+)\, .    
\end{equation}
 The negative part of $\tilde{r}^D_{V}$ is, in general, nonzero, but plays no role in the optimization problem, as that of $\tilde{r}_\rho^D$ vanishes. Therefore, Eq.~(\ref{PrQrrho}) can be recast in the form
 \begin{equation}
     \bar{F}_{[\rho]}(V)
     =
     \tilde{r}_{V}^{+}\cdot \tilde{r}_{\rho}^{+}\, ,
     \label{PrQrrho_plus}
 \end{equation}
involving only the positive parts of the invariant vectors, $\tilde{r}_V^+$, $\tilde{r}_\rho^+$. 

\rema{In the examples we consider below,  $\mathcal{R}^{(s)}$ is plotted embedded in the positive subspace, of dimension $\lfloor s\rfloor$+1 --- accordingly, only the positive parts, $\tilde{r}_V^+$ and $\tilde{r}_{\rho}^+$, of the invariant vectors are shown.} 
We also apply a further rotation and scaling \rema{to obtain a simple parametrization of the locus $\mathcal{R}^{(s)}$}, resulting in coordinates  $(\tilde{x},\tilde{y})$ for $s=1,3/2$ and $(\tilde{x},\tilde{y}, \tilde{z})$ for $s=2,5/2$. We give the explicit transformations from $r_l$ to the above coordinates in Appendix~\ref{App.Exp.Trans}. This choice simplifies the description of the optimal sensors and has to be done on a case-by-case basis. 
\begin{figure}
	\centering
	\includegraphics[scale=0.6]
    {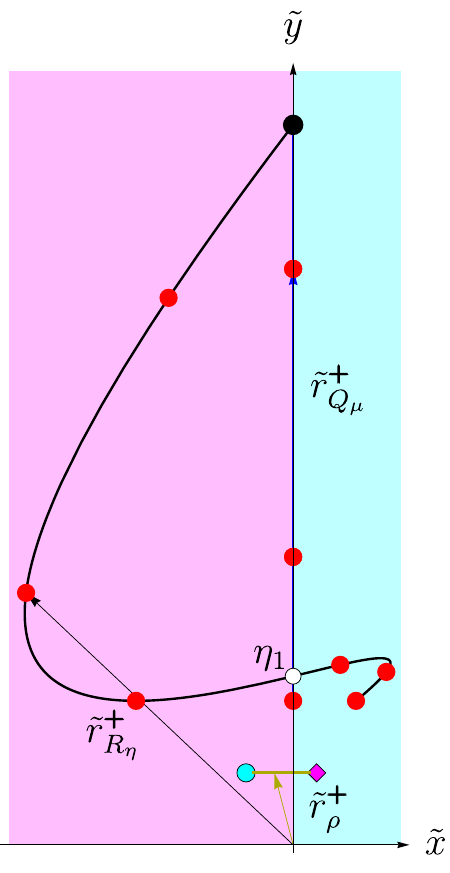}
	\caption{
    Locus $\mathcal{R}^{(3/2)}$ (horizontal mustard yellow segment) and curves corresponding to rotations $R_\eta$ (black) and squeezing $Q_\mu$ (blue) in the spin-$3/2$ $(\tilde{x},\tilde{y})$-plane (\rema{see the caption of Fig.~\ref{Fig.3.spin1locus}} for the visual conventions used). Unlike the $s=1$ case, states with Majorana constellations of different shapes can project to the same point in the $(\tilde{x},\tilde{y})$-plane. 
}
\label{Fig.4.spin32locus}
\end{figure}
\subsubsection{Optimal sensors for rotations and \texorpdfstring{squeezings $S_z^2$}{Lg}.}
\label{SubSe.Optimal}
We consider here the optimal sensors of two standard one-parameter families of finite unitary transformations\rema{:} rotations and one-axis squeezings, defined by
\begin{equation}
    R_{\eta}=e^{-i\eta S_z},
\quad     Q_{\mu}=e^{-i\mu S_z^2},
\end{equation}
respectively. Both are taken with respect to the $z$ axis, which simplifies calculations, but the results are of course valid for arbitrary axes, due to the  $\SUt$ averaging.
For each spin value considered below, we determine which points of the locus $\mathcal{R}^{(s)}$ minimize the average fidelity for these two families.
\subsubsection*{Spin 1}
For $s=1$ the above mentioned linear relations among the components of $r_\rho$ imply it is of the form $r_{\rho}=\left(\frac{1}{3}, r_{1},\frac{2}{3}-r_{1}\right)$, with $r_1$ ranging from 0 (anticoherent state) to 1/2 (coherent state). After rescaling and rotating the positive part of the $\tilde{r}^{D}$- frame to new coordinates ($\tilde{x}$, $\tilde{y}$) (see~\eqref{Eq.Trans.N2}), one ends up with the locus $\mathcal{R}^{(1)}$ being the horizontal segment shown at the bottom of Fig.~\ref{Fig.3.spin1locus}, with the cyan disc on the left end denoting the coherent state, and the magenta square on the right end the anticoherent one. Also drawn in that figure are the $\tilde{r}^{+}$-vectors for rotations (black curve) and squeezings generated by $S_z^2$ (blue curve), for the full range $[0,2\pi]$ of their respective parameters. The plane itself is colored so that if, for a transformation $V$, the vector $\tilde{r}^{+}_{V}$ ends on a specific color, the optimal sensor for $V$ is the similarly colored vertex of $\mathcal{R}^{(1)}$. Note that the rotation curve $\tilde{r}_{R_\eta}^{+}$ crosses from magenta to cyan for the critical value $\eta=\eta_1=\arccos (-2/3)$. For any transformation $V$ such that $\tilde{r}_{V}^{+}$ lies on the dividing line of the plane, so that it is orthogonal to $\mathcal{R}^{(1)}$, all states are equivalent $V$-sensors.
\subsubsection*{Spin 3/2}
For $s=3/2$ the analysis is very similar to that of $s=1$. In this case the linear relations given in Eq. (\ref{reigen}) imply that $r_\rho=\left(\frac{1}{4},r_1,\frac{1}{4},\frac{1}{2}-r_1\right)$. The results are summarized in Fig.~\ref{Fig.4.spin32locus}, with identical conventions in effect. The magenta square at the right end of the segment $\mathcal{R}^{(3/2)}$ stands for the equivalence class of the GHZ state, two states being equivalent if they project to the same point in $r$-space. Note that new coordinates ($\tilde{x}, \tilde{y}$) (see~\eqref{Eq.Trans.N3}), different from those used for $s=1$, are in use here. There is still a critical angle $\eta_1=\arccos\Bigl(\frac{\sqrt{21}-9}{12}\Bigr)$ for rotations, with the novelty that the entire squeezing curve lies on the dividing line, resulting in all spin-3/2 states being equivalent $Q_\mu$-sensors (for any $\mu$!). This echoes the observation that, in the case $s=3/2$, the squeezing-fidelity is the same for all states, for all values of $\mu$. 
\subsubsection*{Spin 2}
\begin{figure} 
	\centering
	\includegraphics[scale=0.59]
    {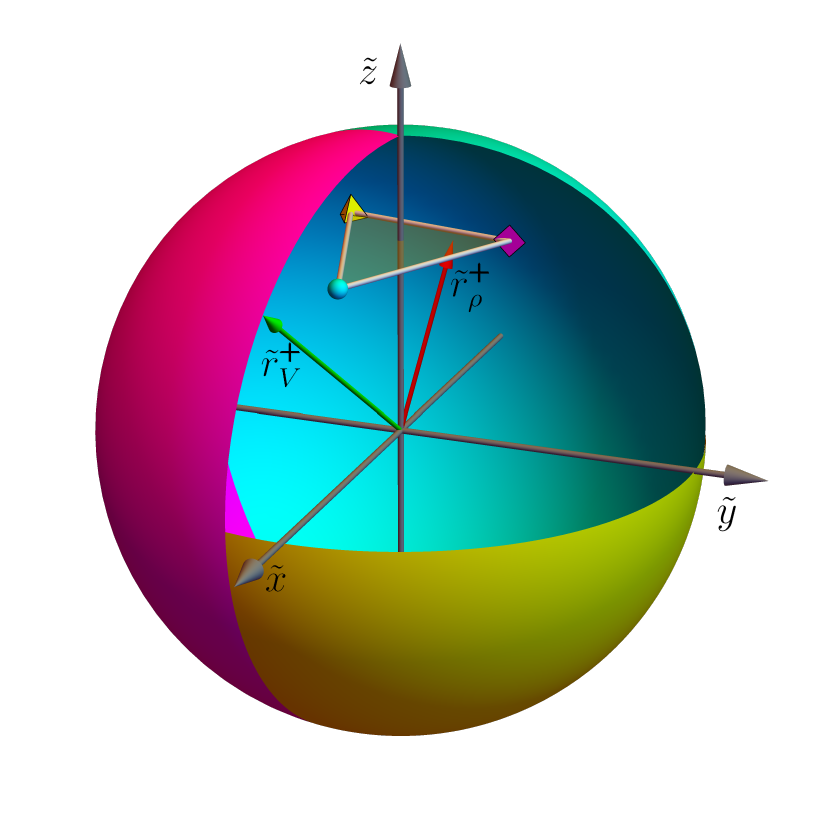}
	\caption{Locus $\mathcal{R}^{(2)}$ of $r_\rho$ for $s=2$. $\mathcal{R}^{(2)}$ is the  triangle in the figure. The vertices correspond to the (classes of) coherent (little cyan sphere), GHZ (magenta square) and tetrahedral (yellow tetrahedron) states. The frame ($\tilde{x}$, $\tilde{y}$, $\tilde{z}$) has been rescaled (with respect to the original $\tilde{r}^{D}_\rho$ referred to) and rotated so that the triangle has constant $\tilde{z}$ component, equal to $1/\sqrt{15}$, and the coherent state vector $\tilde{r}^{+}_\circ$ has $\phi=0$. Also shown are generic vectors $\tilde{r}^{+}_\rho$ (in red) and $\tilde{r}^{+}_{V}$ (in green) for a state $\rho$ and a unitary operator $V$\rema{, respectively} ($\tilde{r}^{+}_{V}$ is normalized to $|\tilde{r}^{+}_{V}|=0.4$ as the optimal state depends only on its direction). The color the vector $\tilde{r}^{+}_{V}$ points to on the sphere indicates the optimal state for detecting the transformation implemented by $V$ (note that the top half of the yellow sector has been removed to provide visual access to the interior). For a given $\tilde{r}^{+}_{V}$, the optimal state is the one minimizing the Euclidean inner product $\tilde{r}_\rho \cdot \tilde{r}_{V}=\tilde{r}^{+}_{\rho}\cdot \tilde{r}^{+}_{V}$.}
    \label{Fig.5.locuss2Plot1}
\end{figure}

For $s=2$, the condition $\lambda r_{\rho} = r_{\rho}$ results in the following form for  $r_{\rho}$,
\begin{equation}
	\label{eq:r_rho_s2}
    r_{\rho}
    =
    \left(
    \frac{1}{5},
    r_{1},
    r_{2},
    \frac{r_{1}-7r_{2}+2}{4},
    \frac{30r_{2}-50r_{1}+12}{40}
    \right)
    \, ,
\end{equation}
so that $\mathcal{R}^{(2)}$ is 2-dimensional (a triangle and its interior), with vertices at 
\begin{alignat}{2}
r_\circ
&=
\Bigl(
\frac{1}{5},\frac{2}{5},\frac{2}{7},\frac{1}{10},\frac{1}{70}
\Bigr)
\quad
&
&\text{coherent state}
\label{s2coh}
\\
r_{\ssssquare}
&=
\Bigl(
\frac{1}{5},0,\frac{2}{7},0,\frac{18}{35}
\Bigr)
&
&\text{GHZ state}
\label{s2GHZ}
\\
r_{\! \tetrahedron}
&=
\Bigl(
\frac{1}{5},0,0,\frac{1}{2},\frac{3}{10}
\Bigr)
&
&\text{tetrahedron state}
\, .
\label{s2tetra}
\end{alignat}
After appropriate rescaling and \rema{rotation}, resulting in new coordinates ($\tilde{x}$, $\tilde{y}$, $\tilde{z}$) (see~\eqref{Eq.Trans.N4}), the triangle ends up perpendicular to the $\tilde{z}$-axis of the (3D) positive subspace of the $\tilde{\lambda}$-matrix, with position vector
\begin{equation}
	\label{eq:r_pos_s2}
    \tilde{r}_{\rho}^{+}
    =
    \left(
   \frac{85 r_{1}+21 r_{2}-12}{42\sqrt{10}},
   \frac{63 r_{2}-25 r_{1}-\rema{8}}{14\sqrt{30}},
   \frac{1}{\sqrt{15}}
    \right)
    \, ,
\end{equation}
see Fig.~\ref{Fig.5.locuss2Plot1}. With this orientation of $\mathcal{R}^{(2)}$ the optimal sensor for a transformation $V$ depends only on the $\phi$ coordinate of $\tilde{r}^{+}_{V}$, with three critical angles given in  Tab.~\ref{tab_phi_s2} --- the coloring of the sphere in Fig.~\ref{Fig.5.locuss2Plot1} changes at these $\phi$-values. 
\begin{table}[t!]
\begin{center}
    \begin{tabular}{|l|c|}
    \hline
    ~Optimal state transition~ & ~Critical angle~ \\
    \hline
    \hline
    ~Tetrahedron to coherent&\rema{$\pi -\tan ^{-1}\left(\frac{5}{\sqrt{3}}\right) \approx 109^{\circ}$}\\
    \hline
    ~Coherent to GHZ&\rema{$\pi +\tan ^{-1}\left(\frac{17}{5 \sqrt{3}}\right) \approx 243^{\circ}$} \\
    \hline
    ~GHZ to tetrahedron&\rema{$2 \pi -\tan ^{-1}\left(\frac{1}{3 \sqrt{3}}\right) \approx 349^{\circ}$}\\
    \hline
    \end{tabular}
    \caption{Critical $\phi$-values for $\mathcal{R}^{(2)}$.}
    \label{tab_phi_s2}
    \end{center}
\end{table}

\begin{figure} 
	\centering
    \includegraphics[scale=0.56]
    {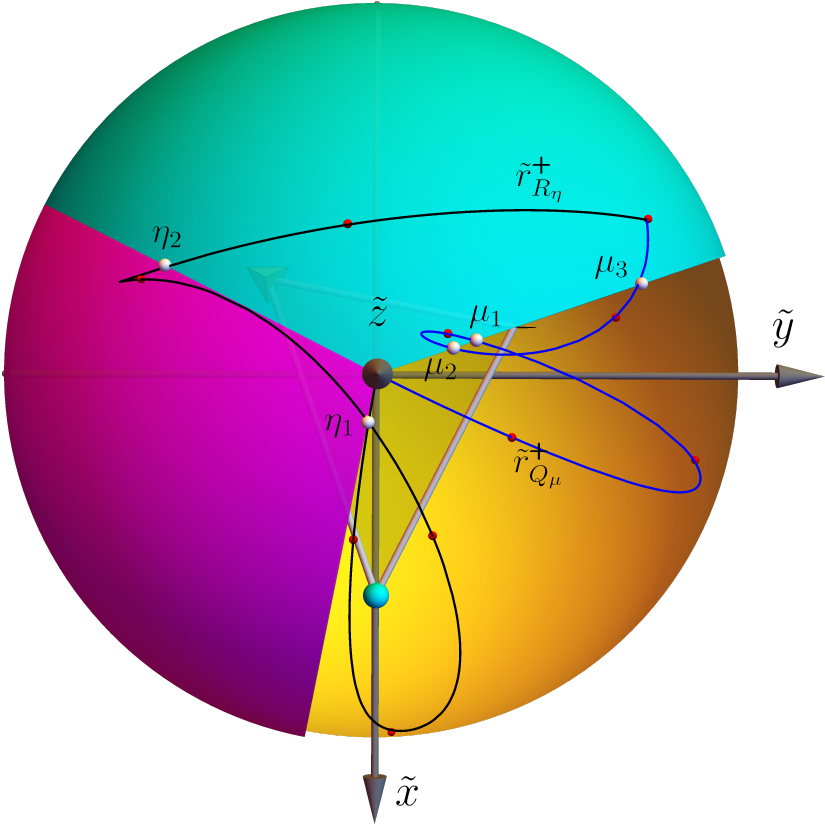}
	\caption{Optimal roto- and squeeze-sensors for $s=2$. The locus $\mathcal{R}^{(2)}$ of Fig.~\ref{Fig.5.locuss2Plot1} is seen here from above. The color transitions are shown in Tab. \ref{tab_phi_s2}. The black curve corresponds to the rotation operators $R_\eta=e^{-i \eta S_z}$, starting, at $\eta=0$, at the identity (at the position of $\tilde{z}$, in the center of the figure), looping in the yellow (tetrahedral) sector and crossing to magenta (GHZ) at $\eta_1=96.47^\circ$, and later on, into cyan (coherent), at $\eta_2=139.95^\circ$. The end point of the curve corresponds to $\eta=\pi$ --- the curve turns back there and retraces itself symmetrically in the $[\pi,2\pi]$-interval. Similarly, the blue curve corresponds to the squeezing operators $Q_\mu=e^{-i \mu S_z^2}$ --- there are three critical points in this case, $\mu_1 \approx 84.82^\circ$, $\mu_2 \approx 120^\circ$, and $\mu_3 \approx 157.45^\circ$. \rema{As} in the case of rotations, the curve retraces itself in the interval $[\pi,2\pi]$. Little white spheres on both curves mark the critical angles $\eta_i$, $\mu_j$, while even smaller red spheres mark $30^\circ$ parameter increments.}
    \label{Fig.6.locuss2Plot2}
\end{figure}

Referring to Fig.~\ref{Fig.6.locuss2Plot2}, the rotation curve $\tilde{r}_{R_\eta}^{+}$ (black curve in the figure) crosses from the tetrahedron (yellow) sector to the GHZ (magenta) sector at $\eta_1 \rema{= 2\arctan(\sqrt{9-2\sqrt{15}})} \approx 96^\circ$ and then to the coherent (cyan)  sector at $\eta_2 \approx 140^\circ$ (the analytical value is given in Ref.~\cite{Mar.Wei.Gir:20}). On the other hand, the squeezing curve $\tilde{r}^{+}_{Q_\mu}$ (in blue) crosses from the tetrahedron sector to the coherent sector at $\mu_1 =\rema{\tan ^{-1}(\sqrt{61+10 \sqrt{37}})}\approx 84.82^{\circ}$, returns to the tetrahedron sector at $\mu_2 = \rema{2\pi/3} =120^{\circ}$, and finally goes back to the coherent sector at $\mu_3 = \rema{\pi -\tan ^{-1}(\sqrt{61-10 \sqrt{37}})} \approx 157.45^{\circ} $. These transitions between sectors double up, symmetrically with respect to $\pi$, as both curves retrace the same locus for $\pi \leq \eta,\mu \leq 2\pi$.

\subsubsection*{Spin 5/2}
\begin{figure}
	\centering
	\includegraphics[width=0.95\linewidth]
    {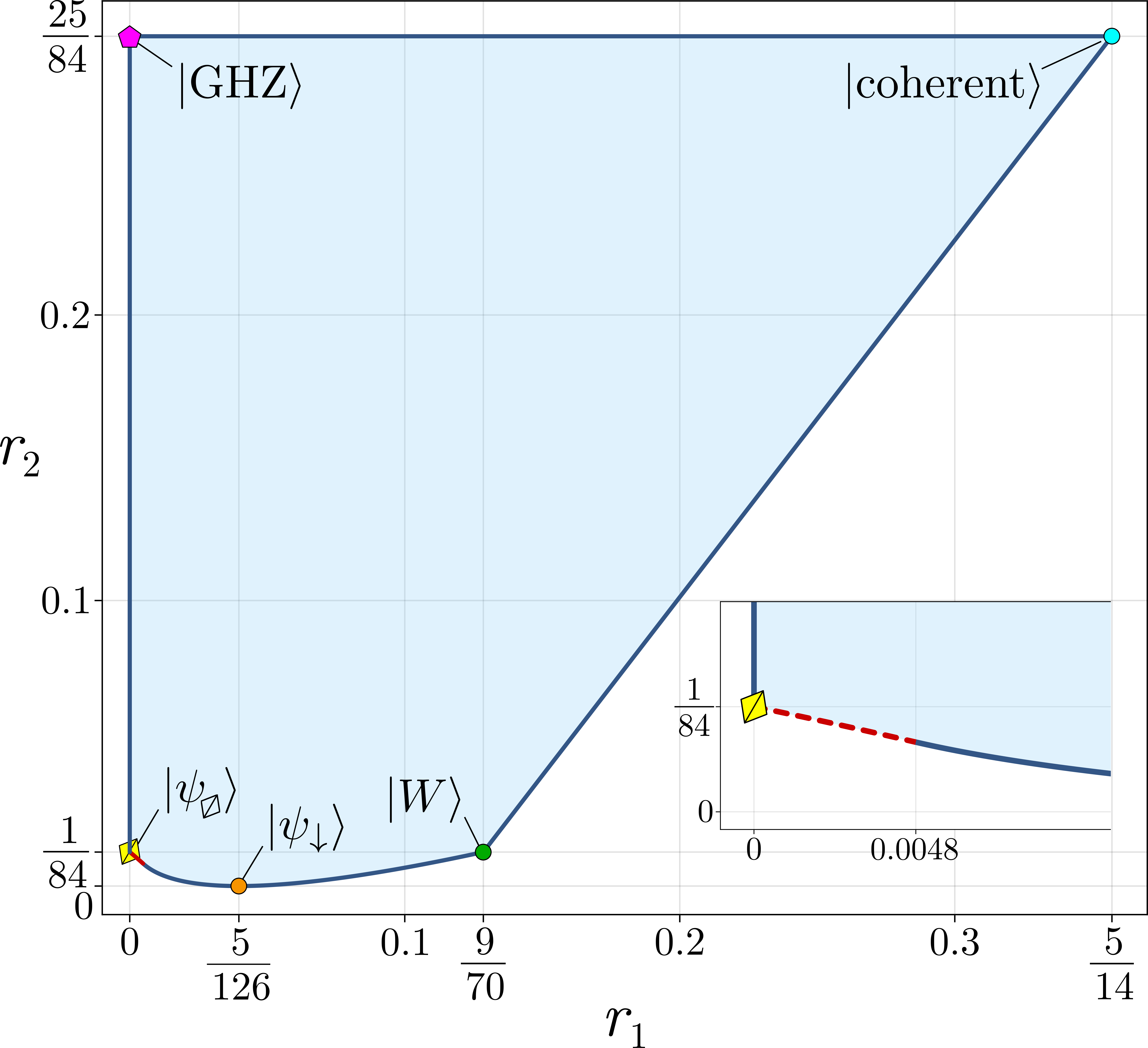}
	\caption{Locus $\mathcal{R}^{(5/2)}$, where the bottom boundary was found only numerically. The vertices represent the coherent state, the pentagonal ($|\mathrm{GHZ}\rangle$) state, the triangular bipyramidal ($|\psi_{\trbyp}\rangle$) state, and the $|W\rangle$ state. Also shown is the state $|\psi_\downarrow\rangle$ with $r_2=0$. Inset: Close-up of the lower left part of the locus. The red dashed curve corresponds to optimal states defined implicitly as a particular real root of a quartic polynomial (see main text). Optimal and worst sensors can only lie on the boundary of $\mathcal{R}^{(5/2)}$.}
	\label{Fig.7.poly_5_2}
\end{figure}

For spin $s=5/2$ the vector  $r_{\rho}$ depends again on two parameters,  $r_{1}$ and $r_{2}$,

\footnotesize{
\begin{equation}
r_\rho\!=\!\left(\frac{1}{6}, r_1,r_2, \frac{7 (r_{1}\!\!-\!2r_{2})}{18}\!+\! \frac{25}{108}, \frac{1}{3}\!-\!r_2, \frac{(14 r_{2}\!-\!25 r_{1})}{18}\!+\!\frac{29}{108}\right).
\end{equation}}
\normalsize
\rema{The locus $\mathcal{R}^{(5/2)}$ is thus two-dimensional. In contrast with the lower spin cases, however, numerical exploration by nonlinear optimization indicates that its projection onto the $(r_1,r_2)$ plane is not a polytope. In particular, the lower part of the boundary (see Fig.~\ref{Fig.7.poly_5_2}), appears to be curved. As a consequence, for certain unitary transformations, the optimal sensor may vary continuously with the transformation parameter over some parameter ranges. The $(r_1,r_2)$ coordinates of the vertices shown in Fig.~\ref{Fig.7.poly_5_2} are given in Table~\ref{tab:vertices_R_5_2}.}
\begin{table}[t]
\renewcommand{\arraystretch}{1.6}
\centering
\begin{tabular}{c c}
\hline
Coordinates & State \\
\hline
\\[-3mm]
$\left(\dfrac{5}{14},\dfrac{25}{84}\right)$
&
$|\mathrm{coherent}\rangle=(1,0,0,0,0,0)$
\\[2mm]
$\left(0,\dfrac{25}{84}\right)$
&
$\displaystyle |\mathrm{GHZ}\rangle=\frac{1}{\sqrt{2}}(1,0,0,0,0,1)$
\\[2mm]
$\left(0,\dfrac{1}{84}\right)$
&
$\displaystyle
|\psi_{\trbyp}\rangle
=\frac{1}{4}\left(0,3,0,\sqrt{2},0,-\sqrt{5}\right)$
\\[2mm]
$\left(\dfrac{9}{70},\dfrac{1}{84}\right)$
&
$\displaystyle |W\rangle=(0,1,0,0,0,0)$
\\[2mm]
$\left(\dfrac{5}{126},0\right)$
&
$\displaystyle |\psi_\downarrow\rangle=\frac{1}{\sqrt{6}}(0,\sqrt{5},0,0,0,1)$
\\[2mm]
\hline
\end{tabular}
\caption{
Vertices and $r_2=0$ point of the projection of $\mathcal{R}^{(5/2)}$ in the $(r_1,r_2)$ plane, highlighted in Fig.~\ref{Fig.7.poly_5_2}.}
\label{tab:vertices_R_5_2}
\end{table}

\begin{figure}
	\centering
	\includegraphics[scale=0.56]
    {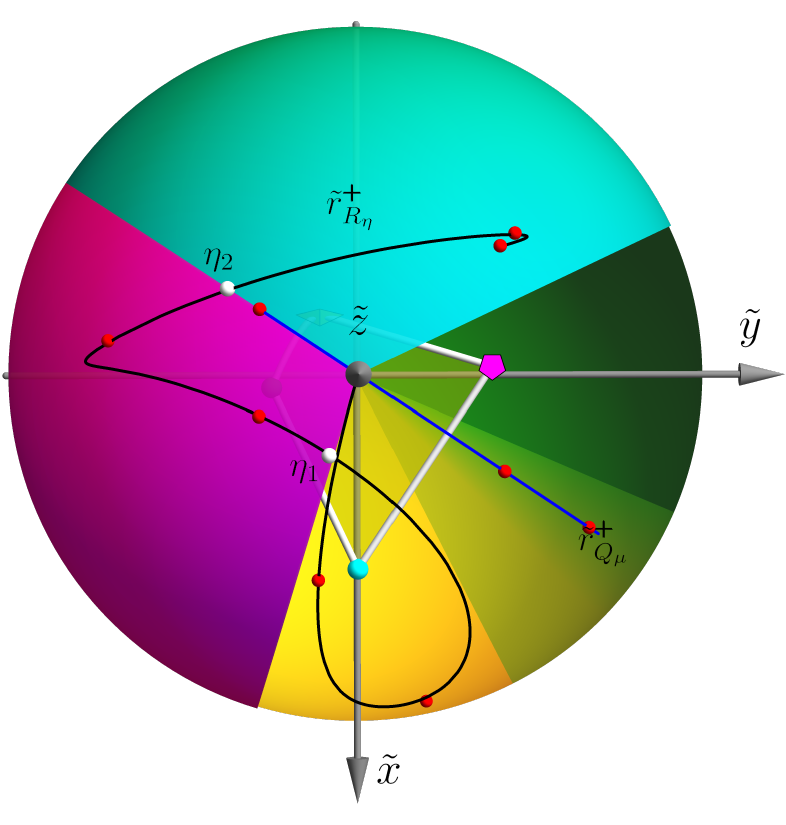}
    \caption{Optimal roto- and squeeze-sensors for $s=\frac{5}{2}$ (refer to Fig.~\ref{Fig.6.locuss2Plot2} for context and visual conventions). The color transitions are given in Tab. \ref{tab_phi_s5_2}. The rotation curve (in black) loops into the yellow (triangular bipyramidal) sector and crosses to magenta (GHZ) at $\eta_1 \approx 86^\circ$, and later on, into cyan (coherent), at $\eta_2\approx 129^\circ$. The end point of the curve corresponds to $\eta=\pi$ --- the curve turns back there and retraces itself symmetrically in the $[\pi,2\pi]$-interval. There are four critical points on the squeezing curve (in blue), $\mu_1 \approx 48^\circ$, $\mu_2 \approx 76^\circ$, $\mu_3 \approx 104^\circ$ and $\mu_{4} \approx 132^{\circ}$, at each of which the curve (which projects as a straight segment when viewed from ``above'') crosses the north pole (identity transformation). Like in the case of rotations, the curve retraces itself in the interval $[\pi,2\pi]$.}
    \label{Fig.8.curves_5_2}
\end{figure}

The lower, curved portion of the boundary of $\mathcal{R}^{(5/2)}$ (in blue, in Fig.~\ref{Fig.7.poly_5_2}) is conjectured to be described by the expression~\cite{GiraudMartinSerranoInProgress}
\begin{equation}\label{r2r1analytical}
r_2(r_1)=
\frac{5 \left(126\, r_1-6 \sqrt{70\, r_1}+5\right)}{1344},
\end{equation}
for
\begin{equation}
0.0048 \approx \frac{22805-9840\sqrt{5}}{166334}
\leq r_1\leq \frac{9}{70} .
\end{equation}
Equation \eqref{r2r1analytical} holds for states of the form $|\psi(\lambda)\rangle=\left(0,\frac{1}{\sqrt{1+\lambda^2}},0,0,0,
\frac{\lambda}{\sqrt{1+\lambda^2}}\right)$ with $0<\lambda\leq 2-\frac{3}{\sqrt{5}}$.
The remaining curved segment (in dashed red) is instead specified implicitly, as a particular real root of a quartic polynomial.
Switching to an appropriate ($\tilde{x}$, $\tilde{y}$, $\tilde{z}$)-frame (see~\eqref{Eq.Trans.N5}), $\mathcal{R}^{(5/2)}$ lies on a constant-$\tilde{z}$ plane, with
\begin{equation}
	\label{eq:r_pos_s5h2}
	\tilde{r}_{\rho}^{+}
    =
    \left(
    \frac{1050 r_1 \! + \! 336 r_2 \! - \! 115}{36 \sqrt{2310}},
    \frac{-\! 210 r_1 \! + \! 336 r_2 \! - \! 25}{36 \sqrt{210}},
    \frac{1}{\sqrt{21}}
    \right) ,
\end{equation}
 which, once more, permits the determination of the optimal sensors of a particular transformation $V$ based only on the $\phi$-coordinate of the corresponding $\tilde{r}_{V}^{+}$ vector --- the coloring of the sphere in Fig.~\ref{Fig.8.curves_5_2} reflects this, with the novelty that a portion of it is now colored with a color gradient. The critical $\phi$-values are shown in Tab. \ref{tab_phi_s5_2}. The spin-$5/2$ case is our first example where we cannot \rema{exactly} characterize the whole set of optimal sensors. Nevertheless, we can still do so for a large family of transformations $V$. Indeed, as long as the corresponding $\tilde{r}^+_V$ vector  has angle $\phi \leq 28^{\circ}$ or $\phi \geq 68^{\circ}$ (see Table~\ref{tab_phi_s5_2}), the optimal sensor is one of the first four states shown in Table~\ref{tab:vertices_R_5_2}.

Referring to Fig.~\ref{Fig.8.curves_5_2}, the rotation curve crosses critical $\phi$-values at $\eta_1 \approx 86^\circ$ and $\eta_2 \approx 129^\circ$.
Thus, the optimal rotosensors are \rema{$|\psi_{\trbyp}\rangle$} for $0\leq \eta \leq 86^{\circ}$, GHZ for $86^{\circ}\leq \eta \leq 129^{\circ}$ and coherent for $129^{\circ}\leq \eta\leq 180^{\circ}$, \rema{in agreement with the results found in Ref.~\cite{Mar.Wei.Gir:20}}. 
\begin{table}[t!]
\begin{center}
    \begin{tabular}{|l|c|}
    \hline
    ~Optimal state transition~ & ~Critical $\phi$-value~ \\
    \hline
    \hline
    ~$|\psi_{\trbyp}\rangle$ to continuous&\rema{$\tan ^{-1}\left(\frac{401}{229 \sqrt{11}}\right) \approx 28^{\circ}$}\\
    \hline
    ~Continuous to $|W\rangle$&\rema{$ \tan ^{-1}\left(\frac{11923}{1487 \sqrt{11}}\right) \approx 68^{\circ}$}\\
    \hline
    ~$|W\rangle$ to coherent&\rema{$\pi -\tan ^{-1}\left(\frac{7}{\sqrt{11}}\right) \approx 115^{\circ}$}\\
    \hline
    ~Coherent to GHZ&\rema{$\pi +\tan ^{-1}\left(\frac{5}{\sqrt{11}}\right) \approx 236^{\circ}$}\\
    \hline
    ~GHZ to $|\psi_{\trbyp}\rangle$&\rema{$2 \pi -\tan ^{-1}\left(\frac{1}{\sqrt{11}}\right) \approx 343^{\circ}$}\\
    \hline
    \end{tabular}
    \caption{Critical $\phi$-values for $\mathcal{R}^{(5/2)}$.}
    \label{tab_phi_s5_2}
\end{center}
\end{table}

The squeezing curve goes through the north pole (identity transformation) for 
\begin{equation}
\begin{aligned}
\mu_1 = & \frac{1}{2} \left[\pi +\tan ^{-1}\left(\tfrac{\sqrt{10 \left(6+\sqrt{15}\right)}}{\sqrt{15}-5}\right)\right] ,
\quad \mu_3 = \pi- \mu_{\rema{2}} ,
\\
\mu_2 = & \frac{1}{2} \left[\pi -\tan ^{-1}\left(\tfrac{\sqrt{10\,(6-\sqrt{15})}}{5+\sqrt{15}}\right)\right] , \quad 
\mu_4 = \pi - \mu_{\rema{1}}
    .
\end{aligned}
\end{equation}

Note that $\tilde{r}_{Q_{\mu}}^{+}$ travels along a critical $\phi$-line for some $\mu$-intervals (see below) and transitions, through the north pole, to the color gradient sector. In detail, the optimal sensors for squeezing are in the color gradient zone for $\mu\in [0,\mu_{1}]\cup[\mu_{2},\mu_{3}] \cup[\mu_{4},\pi]$, and at the coherent-1-AC interface for $\mu\in [\mu_{1},\mu_{2}]\cup [\mu_{3},\mu_{4}]$.

A similar approach works, in principle, for higher spin values. In the cases $s=3$, $7/2$, the positive sector of the $r$-space is four-dimensional, so that the direction of the $\tilde{r}^+_V$ vector lies on a unit 3-sphere. Then, one can still visualize the results by stereographic projection onto the equatorial plane (a copy of $\mathbb{R}^3$), so that instead of the colored sphere in, \emph{e.g.}, Fig.~\ref{Fig.8.curves_5_2}, one would have a colored $\mathbb{R}^3$.
\subsubsection{Spin-2 optimal sensors for uniparametric family of quadratic Hamiltonians}
\label{subsec.Generalized.quadratic}
\begin{figure}[t] 
	\centering
	\includegraphics[scale=0.56]{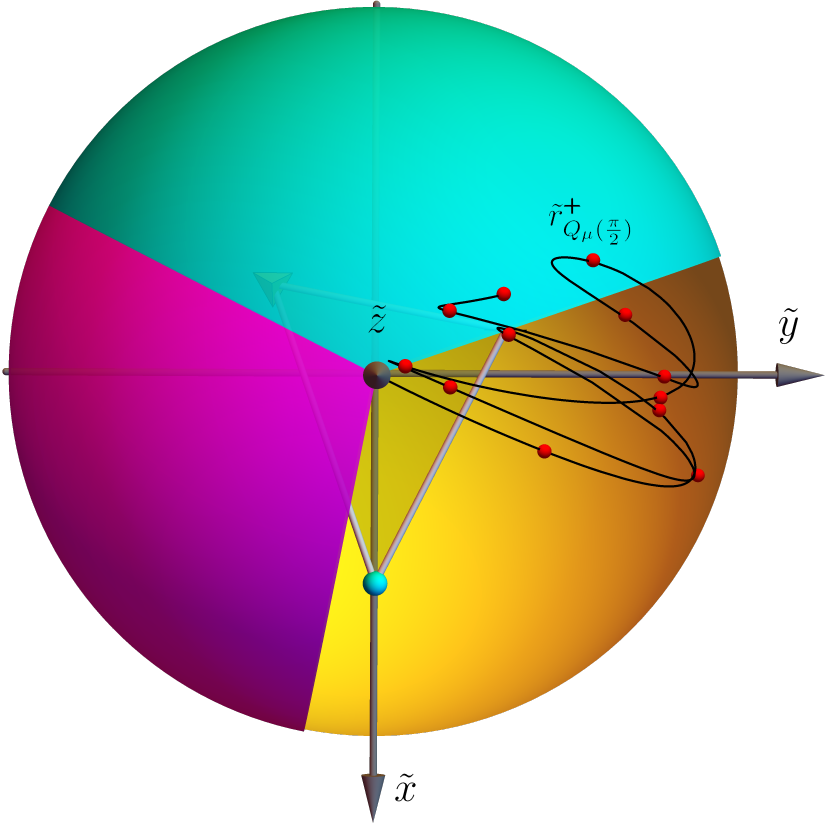}
	\caption{The locus $\mathcal{R}^{(2)}$ of Fig.~\ref{Fig.5.locuss2Plot1} is seen here from above. The color transitions are shown in Tab. \ref{tab_phi_s2}. The black curve corresponds to the squeezing operator $Q_{\mu}(\alpha)=e^{-i \mu H(\alpha)}$ for $\alpha=\pi/2$. The curve starts, at $\mu=0$, at the identity (at the position of $\tilde{z}$, in the center of the figure), looping in the yellow (tetrahedral) sector and crossing to cyan (coherent). The end point of the curve corresponds to $\mu=6\pi$. Little red spheres along the curve mark $90^\circ$ parameter increments.}
    \label{Fig.9.fig_Squeezing}
\end{figure}

We now consider the $\alpha$-family of quadratic traceless Hamiltonians in~\eqref{Eq.Ten.Quadratic}. For illustration, we take $\alpha=\pi/2$, and consider the curve $Q_{\mu}(\pi/2) = e^{-i \mu H(\pi/2)}$, for $s=2$. Note that, \rema{in contrast to the} previous cases, this curve is not periodic in $\mu$ because the eigenvalues of the Hamiltonian are incommensurable  (see Fig.~\ref{Fig.9.fig_Squeezing}). A second distinction, compared to previous cases, is that here one must integrate over the full $\mathrm{SO}(3)$ group to compute the corresponding vector $r_{Q_{\mu}}$ (in the previous cases, a rotational symmetry around an axis allowed the reduction of the integration to one over the sphere). Despite the complicated form of the curve $r_{Q_\mu}$, for any given $\mu$, the color of the point $r_{Q_\mu}$ lands on the sphere in Fig.~\ref{Fig.9.fig_Squeezing} determines the corresponding (class of) optimal sensors. \rema{Visual inspection shows that, in the considered interval of $\mu\in[0,6\pi]$},  there are numerous transitions between the tetrahedral and the coherent sector.
\section{Conclusions}
\label{Sec.Conclusions}
In this work, we studied the problem of identifying optimal quantum sensors (spin-$s$ pure states) for a general unitary transformation, using a fidelity criterion, averaged over all state orientations. 
Our geometric approach reduces the problem to evaluating an inner product between $\mathrm{SU}(2)$-invariant vectors associated with the pure state and the unitary transformation considered. Our main result is a universality principle: there exists a null-measure subset of quantum states (mapped to the boundary of the locus $\mathcal{R}^{(s)}$ of the invariant vectors of all states), among which the optimal (and worst) sensors of any given transformation lie --- the subset itself is independent of the transformation. This result explains in an obvious manner several features of optimal sensors previously discovered, \emph{e.g.}, (i) the same states can be optimal and worst sensors for different transformations (or even the same type of transformation, but for different parameter ranges) (ii) for a given type of transformation, there exist critical values of the associated parameter at which there is a discontinuous change in the corresponding optimal sensor and (iii) for certain transformations, there exists a continuum of optimal sensors.
It is often the case that the optimal sensors found correspond to anticoherent states. In particular, we have demonstrated that anticoherent states play a crucial role as optimal sensors for transformations such as rotations and squeezings. Furthermore, we have highlighted critical transitions in the optimality of sensors at specific values of the transformation parameters, such as rotation angles, which depend on the spin quantum number $s$.

A complete characterization of $\mathcal{R}^{(s)}$, for a particular spin value $s$, permits a similarly complete characterization of rotation-averaged optimal sensors for arbitrary unitary transformations. In this work, such a characterization has been obtained for all spins $s\leq 2$. Our results apply not only to single-spin quantum systems, but also to multi-spin systems in which only the collective spin degree of freedom is used for metrological purposes~\cite{PhysRevLett.124.170401}, as well as to multipartite systems equivalent to a spin-$s$ system, such as multiphoton systems~\cite{Ferretti:24}. For higher spins, the complete characterization of $\mathcal{R}^{(s)}$ remains an open and difficult problem. Nevertheless, some of its boundary components can be known exactly, which allows us to determine the optimal sensors for transformations whose invariant vectors point in the corresponding range of directions. This is the strategy we followed for $s=5/2$: several boundary components of $\mathcal{R}^{(5/2)}$ can be determined exactly, even though the lower boundary is still only partially characterized. A natural direction for future work is therefore to identify exact boundary families of $\mathcal{R}^{(s)}$ for arbitrary spin $s$, especially those with simple scaling properties as $s$ increases. Such results could be useful for understanding asymptotic regimes in quantum metrology.

Any physical system is inevitably subject to noise and decoherence, which convert pure states into mixed states. This regime has been extensively studied in quantum metrology~\cite{PhysRevX.1.021022,PhysRevLett.123.250502,PhysRevA.92.032317}, including in the context of $\SUt$-averaged metrological tasks~\cite{PhysRevA.111.022435}. In future work, we plan to extend our results to mixed states, for which several fidelity measures may be employed~\cite{Liang_2019}.

This work clarifies the link between the shape of the locus $\mathcal{R}^{(s)}$ of invariant vectors of pure states and metrological performance, offering a general framework for quantum sensing. Future work could extend this approach to multipartite systems, and experimental implementations, thereby driving the development of more reliable quantum sensors and advancing quantum metrology.

\section*{Acknowledgments}
The authors thank Uwe R. Fischer for his correspondence. Financial support is acknowledged, in the form of a SECIHTI (Mexico) postdoctoral fellowship (LFAM), a SECIHTI SNI assistantship (AGFD) and the SECIHTI CBF-2025-I-676 and DGAPA-PAPIIT-UNAM IN112224 projects (LFAM, CC, AGFD).
ESE acknowledges support from the postdoctoral fellowship of the IPD-STEMA program of the University of Liège (Belgium). JM and ESE acknowledge the FWO and the F.R.S.-FNRS for their funding as part of the Excellence of Science program (EOS project 40007526).
\appendix
\section{Family of quadratic Hamiltonians}
\label{App.Quadrupolar}
Here, we show that any quadratic Hamiltonian of the form given in~\eqref{Eq.Quadratic.Ham} is equivalent, up to a rotation, to a Hamiltonian of the uniparametric family \eqref{Eq.Ten.Quadratic}. In terms of the polarization tensor operators (see Appendix~\ref{Sub.Polarization}), $H_Q$ has the expression
\begin{equation}
    H_Q = (H_{Q})_{00} T_{00} + \sum_{M=-2}^2  (H_{Q})_{2M}T_{2M} ,
\end{equation}
with 
\begin{equation}
\begin{aligned}    
    (H_{Q})_{00} &= \frac{\sqrt{2s+1}s(s+1)}{3} \left( Q_{xx} + Q_{yy}+Q_{zz} \right) , 
    \\
(H_{Q})_{20} &= - \frac{\sigma(s)}{\sqrt{6}} \left( Q_{xx}+Q_{yy}-2Q_{zz} \right)
\\
(H_{Q})_{21} &= - \sigma(s)\left( Q_{xz} - i Q_{yz} \right)
\\
(H_{Q})_{22} &= \frac{\sigma(s)}{2}  \left( Q_{xx}-Q_{yy} -2 i Q_{xy} \right)
\end{aligned}
\end{equation}  
and $\sigma(s)=(\sqrt{(2s+3)!/(2s-2)!})/(2\sqrt{30})$. Without loss of generality, we can consider a traceless Hamiltonian $(H_{Q})_{00}=0$. Since the $T_{2M}$ operators behave, under rotations,  as the $S_z$-eigenvectors $(s=2,M)$, the vector $((H_{Q})_{22},\dots, (H_{Q})_{2,-2})$ describes a spin-2 state. Then, its Majorana constellation consists of four points. Moreover, $(H_{Q})_{2M}^* =(-1)^M (H_{Q})_{2,-M}$ by the hermiticity of $H_Q$, and consequently, the corresponding Majorana constellation consists of two pairs of antipodal points~\cite{Read2025platonicdynamical}, defining a rectangle inscribed in a great circle. Using an appropriate rotation, the above rectangle  can be oriented such that the four points lie on the equator, and the sides of the rectangle are parallel to the Cartesian axes $x$, $y$. The four complex roots associated with the vertices of the rectangle are then $\zeta_{k}= \pm i e^{\pm i\phi}$ with $\phi\in [0,\pi/4]$. Using the Majorana \rema{bijection~\cite{Chr.Guz.Ser:18}}, we obtain that the \rema{corresponding} normalized state has 
 \rema{
 \begin{equation}
\label{Eq.HQ}
(H_{Q})_{22}=\sqrt{\frac{3}{2(3+\cos^2 (2\phi))}}  , \quad (H_{Q})_{20}=\frac{\cos (2\phi)}{\sqrt{3+\cos^2 (2\phi)}}    
\end{equation}
which are positive numbers, 
and $(H_{Q})_{21}$ is always zero. Therefore, the Hamiltonian $H_Q$ can always be written, up to a rotation, as the linear combination $(H_{Q})_{20} T_{20} + (H_{Q})_{22}(T_{22}+T_{2-2}) = \frac{1}{\sqrt{2}}\Bigl((H_{Q})_{20} \Gamma_{20} + (H_{Q})_{22} \Gamma_{22}\Bigr)$. We consider an extended uniparametric family
$(H_{Q})_{22}=\frac{\sin \alpha}{\sqrt{2}}$ and $(H_{Q})_{20}= \cos \alpha$, with $\alpha \in [0, \pi/2]$, which contains the admissible values in the parametrization~\eqref{Eq.HQ}. Thus, we arrive at 
Eq.~\eqref{Eq.Ten.Quadratic}}.
\section{Polarization tensor basis}
\label{Sub.Polarization}
%
An arbitrary $n \times n$ matrix $A$ may be expanded in the (non-Hermitian) polarization tensor basis 
$\{T^{(s)}_{lm}: l=0,1,\ldots,2s$, $m=l,\ldots,-l\}$~\cite{Var.Mos.Khe:88} (referred to as the $T$-basis henceforth) which satisfies 
$\Tr(T^{(s)}_{lm}T^{(s) \, \dagger}_{l' m'})=\delta_{ll'}\delta_{mm'}$, so that
\begin{equation}
    \label{ATexp}
    A=\sum_{l=0}^{2s} \sum_{m=-l}^l A_{lm} T_{lm}
    \, ,
    \quad A_{lm}=\Tr \big(A T_{lm}^\dagger \big)
    \, ,
\end{equation}
where here, and often in what follows, we omit the superindex $(s)$.
We denote the corresponding vectorization, in this basis, with a subscript $T$, $A \mapsto \ket{A}_T=(A_{00},A_{11},A_{10},\ldots,A_{2s,-2s})^T$. The hermiticity and unit trace conditions of a pure state $\rho$ imply the following conditions on its components in the $T$-basis, respectively,
\begin{equation}
    \label{rhocompTb}
    \rho_{l,-m}=(-1)^m \bar{\rho}_{lm}
    \, ,
    \quad
    \rho_{00}=\frac{1}{\sqrt{n}}
    \, .
\end{equation}
The property $\rho^2=\rho$ imposes additional relations among the $\rho_{lm}$.

The $T_{lm}$'s transform irreducibly under the adjoint action of $\SUt$, 
\begin{equation}
    \label{Tlmirrep}
    D^{(s)}(g) T_{lm} D^{(s)}(g)^\dagger=\sum_{m'=-l}^l D^{(l)}(g)_{m' m} T_{lm'}
    \, ,
\end{equation}
where $D^{(s)}(g)$ is the $(2s+1)$-dimensional irreducible representation ($s$-irrep) of $g \in \SUt$. In some instances, it will prove convenient to work in a Hermitian orthonormal basis of $\mathfrak{u}(n)$, according to the inner product in~\eqref{metricun}. We define one as follows (with $\ell=0,\ldots,2s$)
\begin{equation}
\Gamma_{\ell m} =
\left\{
\begin{alignedat}{2}
    & T_{\ell m} +(-1)^m T_{\ell, -m} \, , & \quad 1&{} \leq m \leq \ell \, , \\
    & -i \left(T_{\ell m} -(-1)^m T_{\ell, -m} \right) \, , & \quad -\ell &{}\leq m \leq -1 \, , \\
    & \sqrt{2} \, T_{\ell 0} \, & &\; m=0.
\end{alignedat}
\right.
\label{Hbndef}
\end{equation}
\section{Properties of the eigenvalues \texorpdfstring{$\lambda$}{Lg}}
\label{App.prop:lambda}
The eigenvalues $\lambda_{ll'}$ of $\mathbb{T}_l$, mentioned in~(\ref{TTLTlp}), are given by 
\begin{equation}
	\label{eq:lambda}
	\lambda_{ll'}=(-1)^{2s+l+l'}(2\ell+1)
    \left\{\begin{array}{ccc}
		s&s&l\\
		s&s& l'
	\end{array}\right\}
    \, ,
\end{equation} 
where $\left\{\begin{array}{ccc}
	j_{1}&j_{2}&j\\
	m_{1}&m_{2}&m
\end{array}\right\}$ are the $6j$-symbols. They satisfy
\begin{align}
	\label{eq:prop_lambda_1}
	(2l'+1)\lambda_{ll'}
    &=
    (2l+1)\lambda_{l'l}
    \, ,
    \\
	\label{eq:prop_lambda_2}
	\displaystyle\sum\limits_{l'=0}^{2s}\lambda_{ll'}\lambda_{l'l''}
    &=
    \delta_{ll''}
    \, ,
    \\
    \label{eq:prop_lambda_3}
    \displaystyle\sum\limits_{l=0}^{2s}\lambda_{ll}
    &=
    2s+1 \, \bigl(\textup{mod}\,2\bigr).
\end{align}
The symmetry property (\ref{eq:prop_lambda_1}) can be deduced by exploiting the symmetry of the $6j$-symbols under the interchange of their rows,
\[
\begin{array}{ll}
	\lambda_{ll'}\!\!\!
    &=
    (-1)^{2s+l+l'}(2\rema{l}+1)
    \left\{
    \begin{array}{ccc}
		s&s&l\\
		s&s&l'
	\end{array}
    \right\}
    \\
	\\
	&=
    \left(
    \frac{2l+1}{2l'+1}
    \right)
    (-1)^{2s+l'+l}(2l'+1)
    \left\{
    \begin{array}{ccc}
		s&s&l'\\
		s&s&l
	\end{array}
    \right\}
    \\
 \\
 &=
 \left(
\frac{2l+1}{2l'+1}
\right) \lambda_{l'l}
\, .
\end{array}
\]

As for the proof of (\ref{eq:prop_lambda_2}), it is sufficient to use the orthogonality property of the $6j$-symbols \cite{Var.Mos.Khe:88},
\begin{equation}
	\label{eq:orthogonality_sixJsymbol}
	\sum\limits_{c}
    (2a+1)(2c+1)
    \left\{
    \begin{array}{ccc}
		j_{1}&j_{2}&a\\
		j_{3}&j&c
	\end{array}
    \right\}
    \left\{
    \begin{array}{ccc}
	j_{1}&j_{2}&b\\
	j_{3}&j&c
\end{array}
\right\}
=
\delta_{ab},
\end{equation}
so that
\[
\begin{array}{ll}
\sum\limits_{l'}
\lambda_{ll'}
\lambda_{l'j}\!\!\!
&=
\sum\limits_{l'}(-1)^{l+j}(2l+1)(2l'+1)
\\
&\phantom{=}
\times
\left\{\begin{array}{ccc}
		s&s&l\\
		s&s&l'
	\end{array}
    \right\}
    \left\{
    \begin{array}{ccc}
		s&s&l'\\
		s&s&j
	\end{array}
	\right\}
    \\
	\\
	&=
    \sum\limits_{l'}
    (-1)^{l+j}(2l+1)(2l'+1)
    \\
 \\
 &\phantom{=}\times
 \left\{
 \begin{array}{ccc}
		s&s&l\\
		s&s&l'
	\end{array}
    \right\}\!
    \left\{
    \begin{array}{ccc}
		s&s&j\\
		s&s&l'
	\end{array}
	\right\}\!
    =
    (-1)^{l+j}\delta_{lj}=\delta_{lj}.
\end{array}
\]

Finally, to prove the property (\ref{eq:prop_lambda_3}), let us first consider the following result for the $6j$-symbols \cite{App:68}:
\footnotesize
\begin{equation}
    \label{eq:prop_six_six}
    \begin{array}{ll}
    \sum\limits_{l_{3}=0}^{2s}(-1)^{j+j_{3}+l_{3}}(2l_{3}+1)
    &\left\{\begin{array}{ccc}
    j_{1}&j_{2}&j_{3}\\
    l_{1}&l_{2}&l_{3}
    \end{array}\right\}\times\\
    \\
    &\left\{\begin{array}{ccc}
    j_{1}&l_{1}&j\\
    j_{2}&l_{2}&l_{3}
    \end{array}\right\}
    =
    \left\{\begin{array}{ccc}
    j_{1}&j_{2}&j_{3}\\
    l_{2}&l_{1}&j
    \end{array}\right\},
    \end{array}
\end{equation}
\normalsize
which allows us to reexpress $\lambda_{ll}$ as
\[\lambda_{ll}=\displaystyle\sum\limits_{J=0}^{2s}(-1)^{J+2s}(2J+1)(2l+1)
\left\{\begin{array}{ccc}
s&s&l\\
s&s&J
\end{array}\right\}
\left\{\begin{array}{ccc}
s&s&l\\
s&s&J
\end{array}\right\}.\]
Subsequently, it suffices to consider the property \cite{Var.Mos.Khe:88}
\begin{equation}
    \label{eq:prop_six_six_2}
    \displaystyle\sum_{X}(2X+1)
    \left\{\begin{array}{ccc}
    a&b&X\\
    c&d&p
    \end{array}\right\}
    \left\{\begin{array}{ccc}
    a&b&X\\
    c&d&q
    \end{array}\right\}
    =
    \displaystyle\frac{\delta_{pq}}{2p+1},
\end{equation}
so that
\[\begin{array}{ll}
\displaystyle\sum\limits_{l=0}^{2s}\lambda_{ll}&
=
\displaystyle\sum\limits_{J=0}^{2s}
(-1)^{J+2s}(2J+1)\times\\
&\phantom{=}
\displaystyle\sum_{l=0}^{2s}(2l+1)
\left\{\begin{array}{ccc}
s&s&l\\
s&s&J
\end{array}\right\}
\left\{\begin{array}{ccc}
s&s&l\\
s&s&J
\end{array}\right\}\\
&=
(-1)^{2s}\displaystyle\sum\limits_{J=0}^{2s}(-1)^{J}=(-1)^{2s}\left(\displaystyle\frac{1+(-1)^{2s}}{2}\right)\\
&=
2s+1 \, (\textup{mod }2).
\end{array}\]
%
%

\section{
On the dimension of \texorpdfstring{$\mathcal{R}^{(s)}$}{Lg}}
\label{App.dimproof}

We prove here that the upper bound
\begin{equation}
\label{Eq.Dim}
\dim \mathcal{R}^{(s)}\leq \bar s,\qquad \bar s=\lfloor s\rfloor ,
\end{equation}
is saturated. Since the linear relations $\lambda r_\rho=r_\rho$, together with the fixed value of $r_0$, leave at most $\bar s$ independent components, it is enough to exhibit one $\bar s$-parameter family of pure states for which the Jacobian $\partial r/\partial \boldsymbol{\alpha}$ of the map from the state space to $\mathds{R}^{\bar{s}}$
\begin{equation}
r(\rho_{\boldsymbol{\alpha}}) \equiv \left( r_1(\rho_{\boldsymbol{\alpha}}),\ldots,r_{\bar s}(\rho_{\boldsymbol{\alpha}}) \right)
\end{equation}
has full rank at one point.

Consider the normalized family
\begin{equation}
\label{eq:dimproof_family}
\ket{\psi_{\bm\alpha}}
=
N(\bm\alpha)
\left(
\ket{\psi_0}
+
\sum_{\mu=0}^{\bar s-1}
\alpha_\mu \ket{s,s-\mu}
\right),
\end{equation}
with
\begin{equation}
    N(\bm\alpha)=
\frac{1}{\sqrt{1+\bm\alpha\cdot \bm\alpha}},
\end{equation}
where $\alpha_\mu\in\mathbb{R}$, and
\begin{equation}
\label{eq:dimproof_psi0}
\ket{\psi_0}
=
\sum_{\nu=\bar s-s}^{s}a_\nu \ket{s,-\nu}
\end{equation}
is normalized. We choose the coefficients $a_\nu>0$; an convenient explicit choice is $a_\nu=1/\sqrt{2s-\bar s+1}$.

Let $r'_l$ denote the quantity $r_l$ in~\eqref{Eq.rl.to.rhol} evaluated on the unnormalized vector inside parentheses in Eq.~\eqref{eq:dimproof_family}. Since
\begin{equation}
r_l(\bm\alpha)=N(\bm\alpha)^4 r'_l(\bm\alpha),
\end{equation}$\partial_{\alpha_\mu}N^4|_{\bm\alpha=0}=0$ and $N(\bm{0})=1$, the Jacobian of the normalized invariants at $\bm\alpha=0$ is the same as the Jacobian of the $r'_l$'s there.

We compute
\begin{equation}\label{eq:dimproof_derivative}
\begin{aligned}
    \left.
\partial_{\alpha_\mu} r'_l
\right|_{\bm\alpha=0}
=
& 2\sum_{m=-l}^{l}
\bra{\psi_0}T_{lm}^{\dagger}\ket{\psi_0}\times\\
&
\left(
\bra{s,s-\mu}T_{lm}\ket{\psi_0}
+
\bra{\psi_0}T_{lm}\ket{s,s-\mu}
\right).
\end{aligned}
\end{equation}
Using the standard selection rule
\begin{equation}
\label{eq:dimproof_selection}
\bra{s,n_1}T_{lm}\ket{s,n_2}
\propto
\delta_{n_1,n_2+m},
\end{equation}
one sees immediately that
\begin{equation}
\left.
\partial_{\alpha_\mu} r'_l
\right|_{\bm\alpha=0}=0
\qquad
\text{if}
\qquad
l<\bar s-\mu .
\end{equation}
Thus, if the rows are ordered as $l=\bar s,\bar s-1,\ldots,1$ and the columns as $\mu=0,1,\ldots,\bar s-1$, the Jacobian is triangular. Its diagonal entries are
\begin{equation}
\label{eq:dimproof_diag}
\left.
\partial_{\alpha_\mu} r'_{\bar s-\mu}
\right|_{\bm\alpha=0}
=
4\,
\bra{\psi_0}
T_{\bar s-\mu,\mu-\bar s}
\ket{\psi_0}\,
a_{\bar s-s}
(-1)^\mu
c^{\bar s-\mu,\mu-\bar s}_{s,s-\bar s;\,s,\mu-s}.
\end{equation}
The Clebsch-Gordan (CG) coefficient in Eq.~\eqref{eq:dimproof_diag} is nonzero. Indeed, using CG symmetries, it is equal  to
\begin{equation}
(-1)^{2s-\bar{s}+\mu} c^{\bar s-\mu,\bar s-\mu}_{s,\bar s-s;\,s,s-\mu},
\end{equation}
which is of the form $c^{cc}_{a\alpha,a\beta}$, with $c=\alpha+\beta$. Its explicit expression is
\begin{equation}
\label{eq:dimproof_cg_nonzero}
c_{a\alpha,a\beta}^{cc}
=
(-1)^{a-\alpha}
\sqrt{
\frac{(2c+1)!(2a-c)!(a+\alpha)!(a+\beta)!}
{(c!)^2(2a+c+1)!(a-\alpha)!(a-\beta)!}
},
\end{equation}
and is nonzero for the values appearing here.

It remains only to show that
\begin{equation}
\bra{\psi_0}
T_{\bar s-\mu,\mu-\bar s}
\ket{\psi_0}
\neq 0
\qquad
(\mu=0,\ldots,\bar s-1).
\end{equation}
With the choice of positive equal coefficients $a_\nu$, this matrix element becomes a sum of positive terms. Explicitly, using the selection rule and Eq.~\eqref{eq:dimproof_cg_nonzero},
\begin{equation}
\label{eq:dimproof_positive_sum}
\bra{\psi_0}
T_{\bar s-\mu,\mu-\bar s}
\ket{\psi_0}
=
\sum_{\nu_1,\nu_2=\bar s-s}^{s}
a_{\nu_1}a_{\nu_2}
(-1)^{s+\nu_2}
c^{\bar s-\mu,\mu-\bar s}_{s,-\nu_2;\,s,\nu_1},
\end{equation}
with
\begin{equation}
c^{\bar s-\mu,\mu-\bar s}_{s,-\nu_2;\,s,\nu_1}= \frac{\delta_{\nu_1-\nu_2,\bar s-\mu}}{(-1)^{s+\nu_2}}
\sqrt{
\frac{(2c+1)!(2s-c)!(s-\nu_2)!(s+\nu_1)!}
{(c!)^2(2s+c+1)!(s+\nu_2)!(s-\nu_1)!}
}
\end{equation}
where $c=\bar s-\mu$. All nonzero summands in Eq.~\eqref{eq:dimproof_positive_sum} are strictly positive, and at least one such summand exists. Hence the matrix element is nonzero.

Therefore every diagonal entry of the triangular Jacobian is nonzero, and
\begin{equation}
\det
\left(
\left.
\frac{\partial r_l}{\partial \alpha_\mu}
\right|_{\bm\alpha=0}
\right)
\neq 0 .
\end{equation}
The map $\bm\alpha\mapsto (r_1,\ldots,r_{\bar s})$ has rank $\bar s$ at $\bm\alpha=0$. Hence $\mathcal{R}^{(s)}$ contains a $\bar s$-dimensional patch. Therefore, by Eq.~\eqref{Eq.Dim}, we conclude that
\begin{equation}
\dim \mathcal{R}^{(s)}=\lfloor s\rfloor .
\end{equation}

\rema{
\section{Explicit transformations}
\label{App.Exp.Trans}
Here, we write the explicit transformations between the invariants $r_{\rho}$ or $r_{V}$ and the variables $(\tilde{x},\tilde{y},\tilde{z})$ of our examples in Section~\ref{SubSec.Finite} for different spin quantum numbers $s$:
\begin{align}
\label{Eq.Trans.N2}
s=1: \quad
&\left\{
\begin{aligned}
\tilde{x}
&=
\frac{3 r_2-5 r_1}{2 \sqrt{30}}\, ,
\\
\tilde{y}
&=
\frac{4 r_0+r_1+r_2}{2 \sqrt{6}}\, .
\end{aligned}
\right.
\\[1em]
\label{Eq.Trans.N3}
s=\frac{3}{2}: \quad
&\left\{
\begin{aligned}
\tilde{x}
&=
\frac{3 r_3-7 r_1}{\sqrt{210}}\, ,
\\
\tilde{y}
&=
\frac{5 r_0+r_1+r_2+r_3}{2 \sqrt{10}}\, .
\end{aligned}
\right.
\\[1em]
\label{Eq.Trans.N4}
s=2: \quad
&\left\{
\begin{aligned}
\tilde{x}
&=
\frac{63 r_1+15 r_2-4 \left(3 r_3+5 r_4\right)}
{42 \sqrt{10}}\, ,
\\
\tilde{y}
&=
\frac{-7 r_1+17 r_2-22 r_3+10 r_4}
{14 \sqrt{30}}\, ,
\\
\tilde{z}
&=
\frac{6 r_0+r_1+r_2+r_3+r_4}{2 \sqrt{15}}\, .
\end{aligned}
\right.
\\[1em]
\label{Eq.Trans.N5}
s=\frac{5}{2}: \quad
&\left\{
\begin{aligned}
\tilde{x}
&=
\frac{
264 r_1+99 r_2-11 \left(r_3+5 r_4\right)-65 r_5
}
{12 \sqrt{2310}}\, ,
\\
\tilde{y}
&=
-\frac{
24 r_1-45 r_2+29 r_3+25 r_4-25 r_5
}
{12 \sqrt{210}}\, ,
\\
\tilde{z}
&=
\frac{7 r_0+r_1+r_2+r_3+r_4+r_5}{2 \sqrt{21}}\, .
\end{aligned}
\right.
\end{align}
}
\bibliographystyle{apsrev4-2}
\bibliography{strings}

\end{document}